\documentclass[fleqn,usenatbib]{mnras}

\usepackage{newtxtext,newtxmath}
\usepackage[T1]{fontenc}

\DeclareRobustCommand{\VAN}[3]{#2}
\let\VANthebibliography\thebibliography
\def\thebibliography{\DeclareRobustCommand{\VAN}[3]{##3}\VANthebibliography}

\usepackage{graphicx}	% Including figure files
\usepackage{amsmath}	% Advanced maths commands
\usepackage{subcaption}
\usepackage{multirow}
\usepackage{physics}
\usepackage{float}
\usepackage{booktabs}
\usepackage{multicol}
\usepackage{threeparttable}
\usepackage{pdflscape}

\title[\texttt{pop-cosmos}: Galaxy size and morphology evolution]{\texttt{pop-cosmos}: Galaxy size evolution across structural and star-formation classifications in COSMOS-Web}

\author[M. N. Tudorache et al.]{Madalina N. Tudorache,$^{1}$\thanks{E-mail: madalina.tudorache@ast.cam.ac.uk}
Hiranya V. Peiris,$^{1,2,3}$
Stephen Thorp,$^{1,3}$
Sinan Deger,$^{1}$
Daniel J. Mortlock,$^{4,5}$
\newauthor
Gurjeet Jagwani,$^{1,6}$
Anik Halder,$^{1}$
Boris Leistedt,$^{4}$
Benedict Van den Bussche$^{1}$
and Joel Leja$^{7,8,9}$
\\
$^{1}${Institute of Astronomy, University of Cambridge, Madingley Road, Cambridge CB3 0HA, UK} \\
$^{2}${Cavendish Laboratory, Department of Physics, University of Cambridge, JJ Thomson Avenue, Cambridge, CB3 0HE, UK} \\
$^{3}${The Oskar Klein Centre, Department of Physics, Stockholm University, AlbaNova University Centre, SE 106 91 Stockholm, Sweden}\\
$^{4}${Astrophysics Group, Imperial College London, Blackett Laboratory, Prince Consort Road, London, SW7 2AZ, UK}\\
$^{5}${Department of Mathematics, Imperial College London, London, SW7 2AZ, UK}\\
$^{6}${Research Computing Services, University of Cambridge, Roger Needham Building, 7 JJ Thomson Ave, Cambridge CB3 0RB, UK}\\
$^{7}${Department of Astronomy \& Astrophysics, The Pennsylvania State University, University Park, PA 16802, USA}\\
$^{8}${Institute for Computational \& Data Sciences, The Pennsylvania State University, University Park, PA 16802, USA}\\
$^{9}${Institute for Gravitation \& the Cosmos, The Pennsylvania State University, University Park, PA 16802, USA}
}

\date{Accepted XXX. Received YYY; in original form ZZZ}

\pubyear{\the\year{}}

\begin{document}
\label{firstpage}
\pagerange{\pageref{firstpage}--\pageref{lastpage}}
\maketitle

\begin{abstract}  
Galaxy sizes are correlated with stellar mass and redshift, as characterised by  size scaling relations. The inferred forms of these scaling relations are sensitive to how galaxies are classified -- either by their star formation activity (e.g.\ specific star-formation rate, sSFR) or by their morphology markers (e.g.\ bulge-to-total ratio, S\'{e}rsic index). 
We combine stellar mass and sSFR estimates from \texttt{pop-cosmos} (a generative model trained on COSMOS2020  \textit{Spitzer} IRAC $\textit{Ch.\,1} <26$) with size and morphology measurements from COSMOS-Web, obtaining $99\,369$ galaxies. By investigating the size--mass and the size--redshift relations, we show that: (i) the sSFR/morphology splits give quantitatively different slopes, intercepts, and intrinsic scatter behaviour; (ii) intrinsic scatter depends on structural morphology but not on sSFR, which constrains the galaxy--halo connection; (iii) the quiescent and bulge-dominated size--mass relations both show double-power law breaks, but at different pivot masses, indicating that quenching and structural transformation occur on different time-scales; (iv) the morphology-dependent trends are only recoverable from space-based imaging. Further,  the quiescent pivot mass $M_{\ast} \sim 10^{10.7}~\mathrm{M}_{\odot}$ coincides with the mass scale at which the active galactic nuclei (AGN, infrared torus) bolometric luminosity fraction peaks in transitioning galaxies, while the bulge-dominated pivot mass $M_{\ast} \sim 10^{11.1}~\mathrm{M}_{\odot}$  coincides with the halo mass at which AGN-driven baryonic redistribution peaks, which may trace the interval over which AGN feedback ramps from quenching onset to structural transformation.

\end{abstract}


%%%%%%%%%%%%%%%%%%%%%%%%%%%%%%%%%%%%%%%%%%%%%%%%%%

%%%%%%%%%%%%%%%%% BODY OF PAPER %%%%%%%%%%%%%%%%%%

\section{Introduction}

Galaxies span a wide range of sizes, structures, and evolutionary stages. Understanding how they assemble their stellar mass and develop their present-day morphologies is central to building a coherent picture of galaxy formation. Since galaxy size reflects the cumulative effects of star formation, mergers, and feedback within dark matter haloes, scaling relations involving size provide an observational probe of these processes. In particular, the size--mass and size--redshift relations link galaxy sizes to stellar mass and cosmic time \citep[e.g.][]{shen03, vanderwel14, lange15, faisst17, mowla19, dimauro19, nedkova21, cutler22}. These relations are typically measured by separating galaxies into star-forming and quiescent populations, although morphological classifications such as the bulge-to-total ratio and S\'{e}rsic index are sometimes treated as equivalent to the star-forming/quiescent division. Where the two have been compared directly (e.g. \citealp{vanderwel14, dimauro19}), the star-forming/quiescent separation has generally relied on rest-frame colour selection and the morphologies on HST or ground-based imaging. Whether the two classification schemes identify the same populations has not yet been examined using SED-based specific star-formation rates (sSFR) together with space-based morphology measurements. SED-based sSFR avoids the contamination affecting $UVJ$ and $NUVrJ$ colour cuts \citep{leja19_uvj, deger25, halder26}.

Previous work has established that star-forming and quenched galaxies, as well as galaxies with different morphologies, follow distinct size--mass relations tracing distinct evolutionary pathways \citep{vandokkum15}. In the conventional picture, star-forming/disc-dominated/late-type galaxies follow a shallower, tighter relation, whereas quenched/bulge-dominated/early-type systems follow a steeper one, especially at the high-mass end \citep{shen03, lange15, mowla19, dimauro19, nedkova21}. This picture extends to redshift evolution. Galaxy sizes decrease systematically with increasing redshift, but the rate of that evolution is not universal. Observed rates of size evolution with redshift \citep[e.g.][]{oesch10, mosleh11, vanderwel14, shibuya15, mowla19, varadaraj24, yang25} fall between the values predicted by different regimes of the \cite{fall80} disc model.
This spread suggests that the rate of size evolution varies across galaxy populations, depending on morphology and evolutionary history.

The ambiguity between star-formation activity and morphology markers matters directly for the interpretation of the size--mass and size--redshift relations, since differences in slope, curvature, intrinsic scatter, and redshift evolution are often used to infer how galaxies formed and transformed. Our aim is therefore not simply to test whether these classifications agree, but to use the way the size--scaling relations change between them as a probe of the distinct physical processes that they each trace. To do so, we combine high-quality estimates of physical properties and morphological indicators to study the relationships between galaxy size, structure, and star formation. We use physical parameter estimates from \citet{thorp25b}, obtained via spectral energy distribution (SED) fitting with the \texttt{pop-cosmos} model \citep{alsing24, thorp25b} and 26-band COSMOS2020 photometry \citep{weaver22}, together with size and morphology measurements from high-resolution JWST imaging in the Cosmic Evolution Survey \citep[COSMOS;][]{scoville07} field, obtained as part of the COSMOS-Web survey \citep{casey23}. The paper is organised as follows. In Section \ref{sec:data}, we present the observational data from COSMOS, and derived data products -- the COSMOS2020 \citep{weaver22}, and COSMOS2025 \citep{shuntov25} catalogues -- that serve as inputs to our analysis. In Section \ref{sec:methods}, we summarise our approach to inferring size--mass and size--redshift relations. We present our results in Section \ref{sec:results}, and discuss their significance in Section \ref{sec:discussion}. We conclude in Section \ref{sec:conclusions}. 

Throughout this work, we assume flat $\Lambda$CDM cosmology with $H_0 = 67.7$ km\,s$^{-1}$\,Mpc$^{-1}$, $\Omega_{\rm m} = 0.31$ and $\Omega_\Lambda = 0.69$ \citep{planck18}. All magnitudes are given in the AB system \citep{oke83, fukugita96}.

\section{Data and Input Catalogues}
\label{sec:data}

Our analysis is based on observations of the COSMOS field, which covers $\sim 2~\deg^{2}$ centred at $10^{\mathrm{h}}00^{\mathrm{m}}28\overset{\mathrm{s}}{.}6$ in RA and $02^{\circ}12'21''$ in dec (J2000 coordinates). We combine two distinct inputs for each galaxy included in our analysis: physical parameters obtained from SED fits to COSMOS2020 photometry (Section~\ref{sec:cosmos2020}) using the \texttt{pop-cosmos} population model (Section~\ref{subsec:pop-cosmos}); and size and morphology markers from the COSMOS-Web survey (Section~\ref{sec:cosmos-web}) as provided in the COSMOS2025 catalogue (Section~\ref{sec:cosmos2025}). We cross match these two input catalogues (Section~\ref{sec:cross-match}) to define our analysis sample. 

\subsection{The COSMOS2020 catalogue}
\label{sec:cosmos2020}
The COSMOS2020 catalogue \citep{weaver22} is built from pre-\textit{JWST} data in the COSMOS field, including: deep \textit{Hubble Space Telescope} (\textit{HST}) imaging \citep{koekemoer07}; extensive ground-based imaging from Hyper Suprime-Cam (HSC; \citealp{taniguchi07, taniguchi15, aihara22}), UltraVISTA \citep{mccracken12}, and Canada-France-Hawaii Telescope's (CFHT) MegaCam \citep{sawicki19}; and \textit{Spitzer} mid-infrared data \citep{ashby13, ashby15, ashby18, moneti22, mcpartland25}. Photometry is based on the \textsc{Farmer} profile-fitting routine \citep{weaver23_farmer}. The catalogue encompasses a wider area than COSMOS-Web, with a `combined' mask of $1.27$~deg$^2$. We work with the \textit{Spitzer} IRAC $\textit{Ch.\,1}<26$ sub-sample of COSMOS2020 that was analysed by \citet{thorp25b} and \citet{deger25}, consisting of 423,262 galaxies.

\subsection{Physical parameters from the \texttt{pop-cosmos} model}
\label{subsec:pop-cosmos}

In order to estimate the physical parameters required for the analysis, we use \texttt{pop-cosmos} \citep{alsing24, thorp25b}, which is a population-level model for galaxies that is learned directly from deep 26-band photometry from COSMOS2020 \citep{weaver22}. Here, the galaxy population is represented as a joint distribution over redshift and intrinsic physical quantities (a total of 16 parameters), such that the model captures realistic correlations and redshift evolution while remaining tied to a physically interpretable SED parametrisation.  

\begin{figure*} 
      \centering
      \captionsetup{justification=centering}
      \includegraphics[width=1\linewidth]{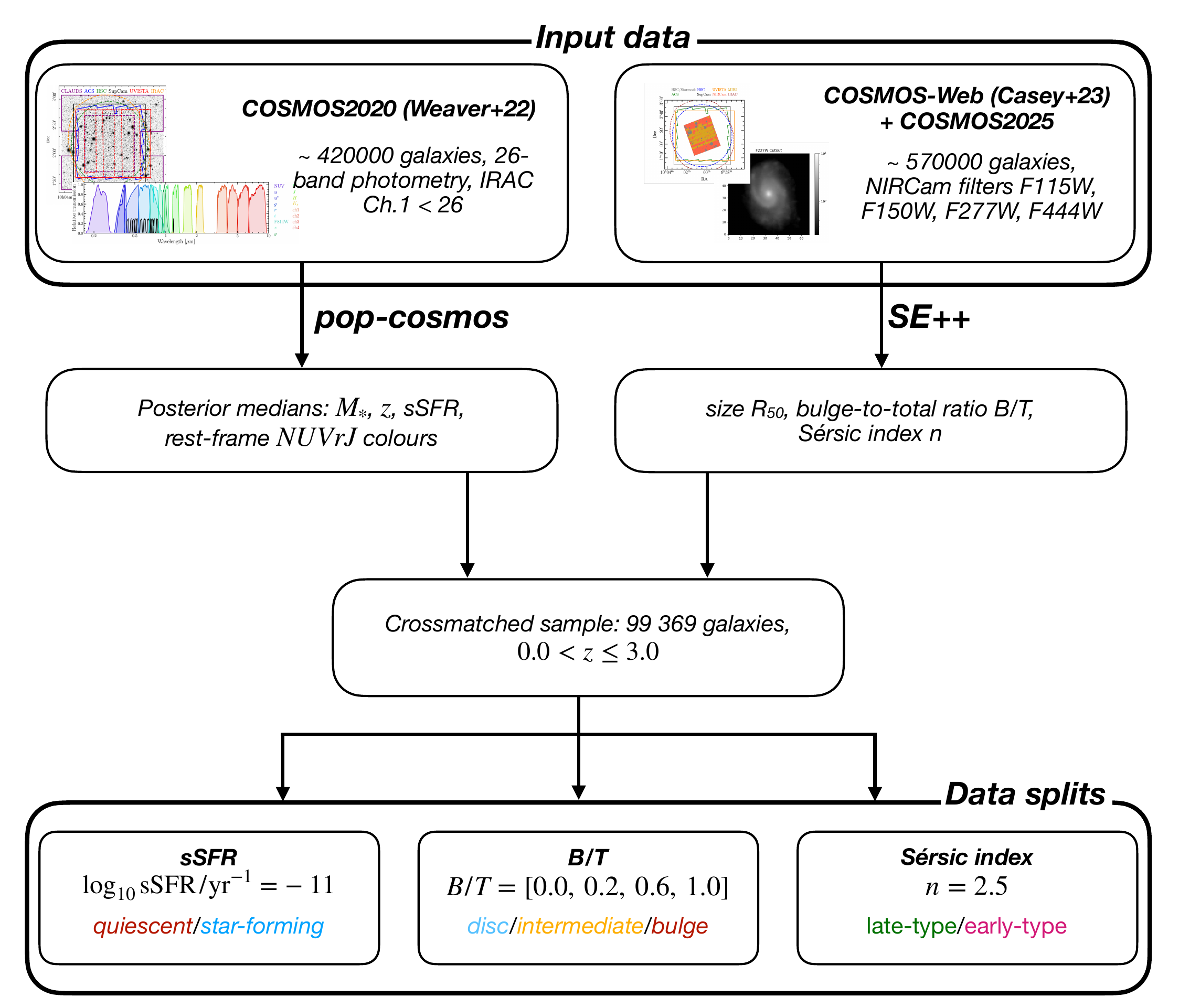} 
      \caption{Schematic of the analysis workflow. Physical parameters (posterior medians of stellar mass $M_{\ast}$, redshift $z$, sSFR, and rest-frame $NUVrJ$ colours) are inferred with \texttt{pop-cosmos} from the deep 26-band COSMOS2020 photometry (\citealt{weaver22}; $\sim 420\,000$ galaxies, IRAC \textit{Ch.1} $< 26$), while sizes and morphology (half-light radius $R_{50}$, bulge-to-total ratio B/T, and S\'{e}rsic index $n$) are measured with \texttt{SourceXtractor++} (SE++) from the COSMOS-Web NIRCam imaging \citep{casey23} as provided in the COSMOS2025 catalogue (\citealt{shuntov25}, $\sim 570\,000$ galaxies). The complementary catalogues are cross-matched into a sample of $99\,369$ galaxies at $0.0 < z \leq 3.0$, which is split by sSFR, B/T, and S\'{e}rsic index as indicated.}
      \label{fig:workflow}
\end{figure*}

The representation of SEDs is based on stellar population synthesis \citep[SPS; for a review see][]{conroy13, iyer25}, as implemented in the Flexible Stellar Population Synthesis (FSPS; \citealp{conroy09, conroy10a, conroy10b}) and \textsc{Prospector} \citep{johnson21} libraries. The specific SPS parametrisation follows a \textsc{Prospector}-$\alpha$-style model \citep{leja17, leja18, leja19_sfh, leja19}. The full details of the parametrisation are described in \citet{alsing24}, \citet{thorp24}, \citet{thorp25b}, and \citet{deger25}, with the 16 SPS parameters listed in the top half of Table \ref{tab:sps-params}. The population distribution in SPS space is learned using a score-based diffusion model \citep{song21}, which can represent complex, high-dimensional densities without imposing restrictive functional forms. The model is calibrated on a large COSMOS2020 training sample: $423\,262$ MIR-selected galaxies with \textit{Spitzer} IRAC $\textit{Ch.\,1} < 26$.

\begin{table}
\centering
\caption{List of SPS parameters (top) and derived parameters (bottom) from the \texttt{pop-cosmos} galaxy population model.}
\label{tab:sps-params}
\begin{tabular}{ll}
\toprule
symbol / unit & definition \\
\midrule
$\log_{10}(M_{\ast,\mathrm{form}}/\mathrm{M}_{\odot})$ & base–10 logarithm of stellar mass formed \\
$\log_{10}(Z/\mathrm{Z}_{\odot})$ & base–10 logarithm of stellar metallicity \\
$\Delta \log_{10}(\mathrm{SFR})_{\{2:7\}}$ & ratios of SFR between neighbouring SFH bins \\
$\tau_{2}$ & diffuse dust optical depth \\
$n$ & index for diffuse dust attenuation law \\
$\tau_{1}$ & birth cloud dust optical depth  \\
$\ln(f_{\mathrm{AGN}})$ & natural logarithm of AGN luminosity fraction  \\
$\ln(\tau_{\mathrm{AGN}})$ & natural logarithm of AGN torus optical depth  \\
$\log_{10}(Z_{\mathrm{gas}}/\mathrm{Z}_{\odot})$ & base–10 logarithm of gas-phase metallicity  \\
$\log_{10}(U_{\mathrm{gas}})$ & base–10 logarithm of gas ionisation \\
$z$ & redshift  \\
\midrule
$t_{\mathrm{age}}/\mathrm{Gyr}$ & mass-weighted age \\
$m_{\ast}$ & base–10 logarithm of stellar mass remaining  \\
$\log_{10}(\mathrm{SFR}/\mathrm{M}_{\odot}\,\mathrm{yr}^{-1})$ & logarithm of SFR  \\
$\log_{10}(\mathrm{sSFR}/\mathrm{yr}^{-1})$ & logarithm of specific SFR \\
\bottomrule
\end{tabular}
\end{table}

\begin{figure*} 
    \begin{subfigure}[b]{1\columnwidth}
      \centering
      % include first image
      \captionsetup{justification=centering}
      \includegraphics[width=1\linewidth]{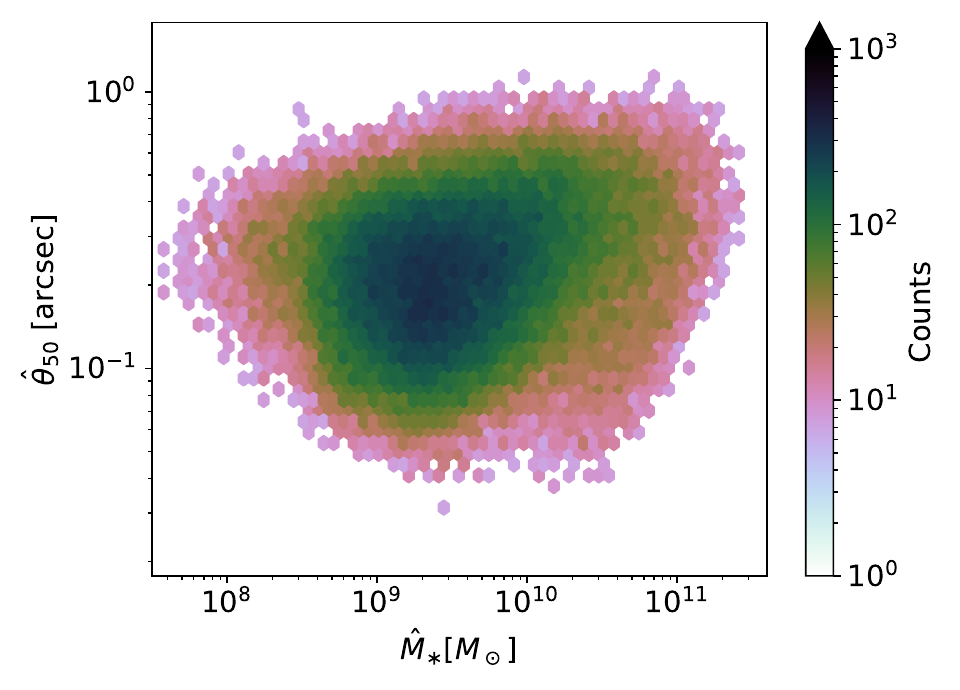} 
      \phantomsubcaption
      \label{subfig:ms-r50-count-cw-arcsec}
    \end{subfigure}
    \begin{subfigure}[b]{1\columnwidth}
        \centering
        \captionsetup{justification=centering}
        \includegraphics[width=1\linewidth]{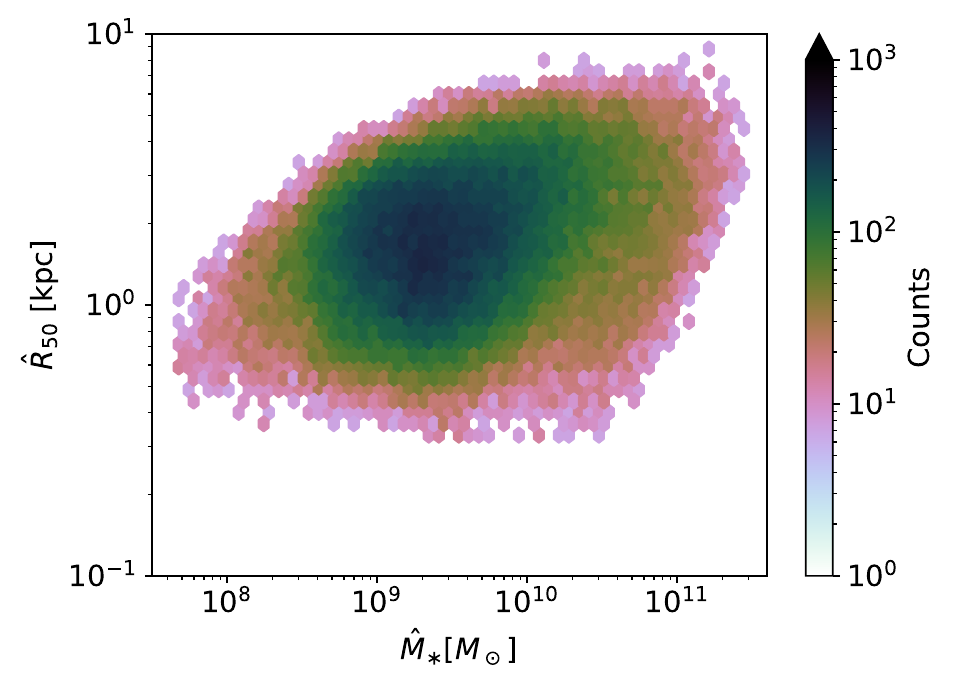}
        \phantomsubcaption
        \label{subfig:ms-r50-count-cw}
    \end{subfigure}
    \caption{The joint distribution of \texttt{pop-cosmos} inferred stellar mass, $\hat{M}_{\ast}$, and measured angular half-light radius $\hat{\theta}_{50}$ (\textit{left}) and physical half-light radius $\hat{R}_{50}$ (\textit{right}) for the crossmatched COSMOS2025 galaxies.}
      \label{fig:ms-r50-cw}
\end{figure*}

The trained \texttt{pop-cosmos} diffusion model can be used as a prior over SPS parameters in downstream SED fitting \citep{thorp24, thorp25b, halder26}. In this work, we use an updated version of the SPS parameter posteriors originally published by \citet{thorp25b} for the sample of COSMOS2020 galaxies described in Section \ref{sec:cosmos2020}. This gives us access to posteriors over the 16 primary parameters listed in the upper part of Table~\ref{tab:sps-params}, as well as the derived parameters listed in the lower part. We additionally compute rest-frame absolute magnitudes in the \textit{GALEX} \textit{NUV}, HSC \textit{r}, and UltraVISTA \textit{J} bands, using the \textsc{Speculator} \citep{alsing20} rest-frame emulators described in \citet{deger25}. These rest-frame magnitudes are used to construct $NUVrJ$ diagrams in Section \ref{sec:discussion}.  Throughout this work, we use parameter estimates and uncertainties based on the \texttt{pop-cosmos} SED fits: posterior median, $\hat{m}_{\ast} \equiv \log_{10}(\hat{M}_{\ast}~/~\mathrm{M}_{\odot})$, and standard deviation, $\sigma_{\hat{m_{\ast}}}$, of logarithmic stellar mass; posterior median, $\hat{z}$, and standard deviation, $\sigma_{\hat{z}}$, of redshift; posterior median sSFR; and posterior median rest-frame $NUV-r$ and $r-J$ colours. We reserve un-hatted symbols for the corresponding theoretical or population-level quantities appearing in the functional forms.

We apply the \citet{thorp25b} stellar mass completeness limit -- based on the estimated peak of the stellar mass function in redshift bins -- to the COSMOS2020 galaxies, based on $\hat{z}$ and $\hat{m}_*$, yielding a sample of $342\,166$ objects. In Appendix \ref{sec:uncertainties}, we examine the typical uncertainties on the \texttt{pop-cosmos} SPS parameter estimates, finding these to be small. The median $z$, $m_{\ast}$, and $\log_{10}(\mathrm{sSFR}/\mathrm{yr}^{-1})$ uncertainties (posterior standard deviations) are $0.09$, $0.12$~dex and $0.41$~dex, respectively. The mass and redshift uncertainties are small enough to be negligible for our subsequent analyses, although we propagate them forward in our inference. Although the typical sSFR uncertainties are higher, the confidence in assigning galaxies as star-forming or quiescent (based on a cut at $\mathrm{sSFR}=10^{-11}\,\text{yr}^{-1}$) is high in most cases (see Appendix~\ref{sec:ssfr_reliability}).

\subsection{The COSMOS-Web survey} 
\label{sec:cosmos-web}
COSMOS-Web is a 255-hour \textit{JWST} Cycle 1 Treasury program (GO 1727; PI: Casey \& Kartaltepe; \citealp{casey23}). It is a wide imaging survey covering $0.54~\mathrm{deg}^2$ in four Near Infrared Camera (NIRCam) filters (F115W, F150W, F277W and F444W), and $0.2~\mathrm{deg}^2$ in the Mid-Infrared Instrument (MIRI) F770W filter. The NIRCam imaging \citep{franco26} reaches $5\sigma$ depth (for $0.15''$ apertures) of $26.6$--$27.3$~mag in F115W, $26.9$--$27.7$~mag in F150W, $27.5$--$28.2$~mag in F277W, and $27.5$--$28.2$~mag in F444W. The MIRI data \citep{harish25} achieve $5\sigma$ depth of $25.33$--$25.98$~mag (for $0.3''$ apertures). The survey footprint forms a roughly square mosaic measuring $46'\times46'$, divided into 152 individual visits. 

\subsection{The COSMOS2025 catalogue}
\label{sec:cosmos2025}

We obtain size and morphology markers (S\'{e}rsic indices, $\hat{n}$, angular half-light radii, $\hat{\theta}_{50}$ and bulge-to-total ratios B/T) from the COSMOS2025 catalogue \citep{shuntov25}, which is built from the COSMOS-Web \textit{JWST} imaging, together with the available \textit{HST}, \textit{Spitzer}, and ground-based photometry in the COSMOS field. It provides homogeneous multi-band photometry from the optical to the mid-IR in 37 photometric bands, measured using \textsc{SourceXtractor++} \citep{bertin96, bertin20, kummel20}, for a parent sample of $784\,016$ galaxies.

Before crossmatching the COSMOS2025 to the COSMOS2020 catalogue, we apply the following quality cuts to ensure reliable measurements:
\begin{itemize}
    \item removing the hot pixels;
    \item removing objects where the space-based measurements were inconsistent with the ground-based measurement (see section 3.7.2 in \citealp{shuntov25});
    \item removing objects that are only detected in one NIRCam band;
    \item removing objects with small radii -- radius~$< 1.3\times10^{-7}$;
    \item removing objects where the flux ratio between the two smallest apertures ($0.2''$ and $0.3''$ diameter) is below $1.6$ for F277W and below $1.7$ for F444W.
\end{itemize}
All these requirements are implemented by choosing \textsc{warn\_flag $= 0$} in the catalogue, to ensure the most secure sources. This cuts the original catalogue from $784\,016$ objects to $691\,285$ objects. We then apply extra cuts to remove stars and blended sources, by requiring that \textsc{flag\_star $= 0$} (reduces the sample to $606\,792$ objects), \textsc{type $= 0$} (reduces the sample to $581\,356$ objects) and \textsc{flag\_blend $= 0$}, which brings the clean sample to $570\,881$ objects. 

\begin{figure*} 
    \centering
    \captionsetup{justification=centering}
    \includegraphics[width=1\linewidth]{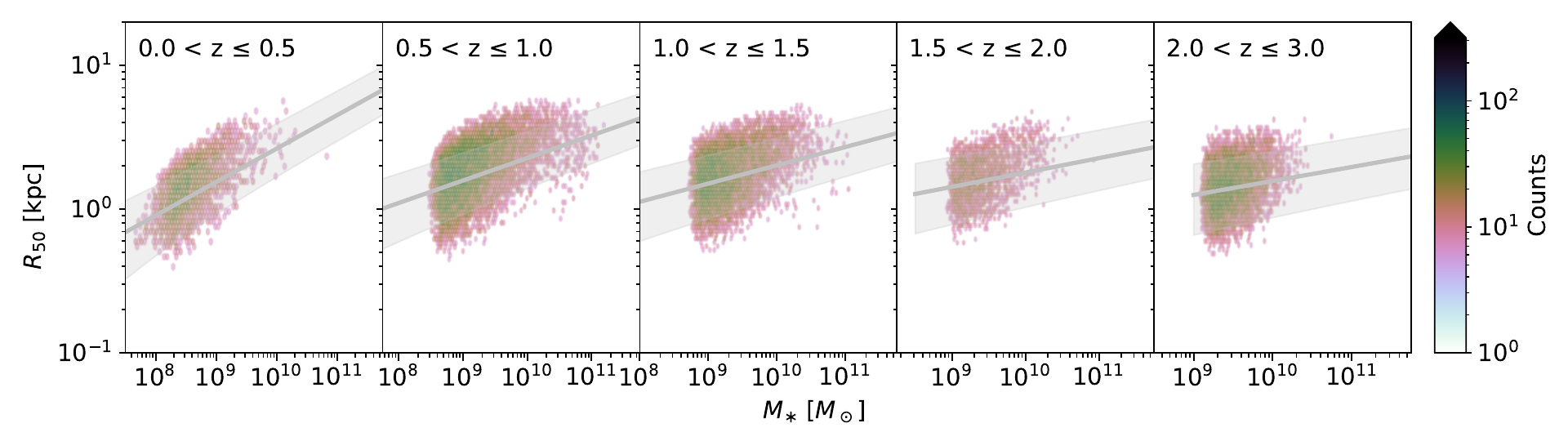} 
    \caption{Size--mass relation in five different redshift bins in the spatially crossmatched COSMOS-Web sample. Cells are shaded by number of galaxies and are only shaded when they contain more than 7 galaxies. The single-power law fit to the size--mass relation is shown by the grey line. The shaded area shows the intrinsic scatter combined with the credible interval (16th and 84th percentile) on the median fit.}
    \label{fig:ms-r50-z-bins-fit}
\end{figure*}

The primary size measurement used is the half-light radius from a S\'{e}rsic profile fit \citep{sersic63, sersic68}. The S\'{e}rsic profile parametrises the surface brightness distribution by the angular half-light radius $\theta_{50}$ and a shape index $n$, where $n=0.5$ corresponds to a Gaussian; $n=1$ is exponential (characteristic of disc galaxies); and $n=4$ recovers the \cite{devaucouleurs48} $\theta^{1/4}$ light profile (characteristic of elliptical galaxies). To estimate the physical half-light radius, $\hat{R}_{50}$, we multiply the angular size by the angular diameter distance corresponding to the posterior median photometric redshift from \texttt{pop-cosmos}, $d_A(z)$. We propagate both the angular size and photometric redshift uncertainties into $\sigma_{\hat{R}_{50}}$, such that
\begin{equation}
    \sigma^2_{\hat{R}_{50}} = d^2_{A}(z)\sigma^2_{\hat{\theta}_{50}} + \hat{\theta}^2_{50}\bigg(\frac{\dd{d_A(z)}}{\dd{z}}\bigg)^2\sigma^2_{\hat{z}}.
\end{equation}

The measured S\'{e}rsic indices, $\hat{n}$, and angular half-light radii, $\hat{\theta}_{50}$, provided in the COSMOS2025 catalogue are calculated by using the four NIRCam bands with \textsc{SourceXtractor++}. They are derived from a joint fit across the four NIRCam bands, and they are not allowed to vary with wavelength. These parameters therefore constitute effective averages of all fitted bands, weighted by a corresponding weight map \citep{shuntov25}. Therefore, the structural parameters correspond to the averaged morphology over the $1$--$5$~\textmu m wavelength range. While this gives better-constrained parameters than single-band fitting, it means that the effective rest-frame wavelength varies with redshift. For the redshift range of this work, the effective rest-frame wavelength probed by the joint fit shifts from approximately $1.1$--$4.4$~\textmu m at $z \approx 0$ (rest-frame near-infrared) to approximately $0.3$--$1.1$~\textmu m at $z \sim 3$ (rest-frame near-UV to near-infrared). The measurement therefore traces progressively bluer rest-frame light with increasing redshift, which is relevant when comparing with studies that measure sizes at a fixed rest-frame wavelength (e.g. $0.5$~\textmu m). The full $1$--$5$~\textmu m NIRCam wavelength coverage means that, in practice, the effective rest-frame wavelength for any given galaxy is pulled toward the highest signal-to-noise bands and will lie somewhere within these intervals\footnote{We note that half-light radii measured from NIRCam rest-frame near-infrared imaging closely trace stellar half-mass radii, with \cite{vanderWel24} finding offsets between $0.01$--$0.07$~dex across all galaxy types and redshifts out to $z \sim 2.5$.}.

As a second morphology indicator, we use the bulge-to-total ratio (B/T) from the COSMOS2025 catalogue which captures structural information complementary to the single-S\'{e}rsic quantities. B/T is computed by fitting a disc+bulge model to each galaxy -- equivalent to fitting a sum of two S\'{e}rsic profiles with fixed $n=1$ and $n=4$, respectively. Unlike the single-S\'{e}rsic quantities, B/T is measured independently in each band. We use the F115W band to estimate the B/T ratio for galaxies at $0.0 < z < 2.0$ and the F150W band for galaxies at $2.0 \leq z \leq 3.0$. This ensures that B/T traces approximately rest-frame optical morphology across the redshift range of interest (see figure~2 in \citealt{yang25}). COSMOS2025 provides this decomposition for the ground-based HSC bands as well as for NIRCam; we use the HSC $i$-band value (B/T HSC-I) in Section \ref{subsec:interchange} as a seeing-limited comparison to the space-based measurement. To ensure there is no discontinuity between the values of the space-based B/T for the two filters at $z = 2.0$, we inspect the rolling median of B/T over redshift, finding no offset at this redshift. Beyond the rolling median, we verified that the full B/T distribution is continuous across the band switch: for galaxies near $z = 2$ (in the redshift interval $1.8 < z < 2.2$), the per-object difference between the F115W and F150W B/T has median $0.00$, with normalised median absolute deviation (NMAD) $\sim 0.04$. This confirms that the switch introduces no offset in either the location or the width of the B/T distribution.

The median values of the uncertainties of the measured B/T and S\'{e}rsic index $\hat{n}$ are $0.03$ and $0.06$, respectively. As the uncertainties are small, the confidence in assigning galaxies in different morphological categories (based on cuts described in Section \ref{subsec:classification}) is high in most cases (see Appendix~\ref{sec:ssfr_reliability}).

The COSMOS2025-derived quantities used in the rest of this work are therefore the estimated physical radius $\hat{R}_{50}$ and its associated uncertainty $\sigma_{\hat{R}_{50}}$, the fitted S\'{e}rsic indices $\hat{n}$ and the estimated bulge-to-total ratios, B/T, for the F115W and F150W filters.

\subsection{Crossmatched analysis sample}
\label{sec:cross-match}
We spatially crossmatch the COSMOS2025 sample from Section~\ref{sec:cosmos2025} to the mass-complete COSMOS2020 sample, described in Section \ref{sec:cosmos2020}, using a $0.5''$ aperture. The result is a catalogue of $111\,847$ galaxies. In Appendix \ref{sec:completeness}, we verify that the crossmatch between COSMOS2020 and COSMOS2025 does not fail in a mass- or redshift-dependent way -- i.e.\ the combined catalogue has no additional selection effect introduced. We calculate the fraction of COSMOS2020 galaxies in bins of $\hat{z}$ and $\hat{m}_{\ast}$ that do not have a successful crossmatch in our COSMOS2025 sample, correcting for the different areas observed by the two catalogues. The crossmatch success rate ranges from $80$--$100$~per cent across the mass–redshift plane, indicating some incompleteness, but we find no systematic dependence on stellar mass or redshift above the completeness limit.

\begin{figure*} 
    \centering
    \captionsetup{justification=centering}
    \includegraphics[width=1.1\linewidth]{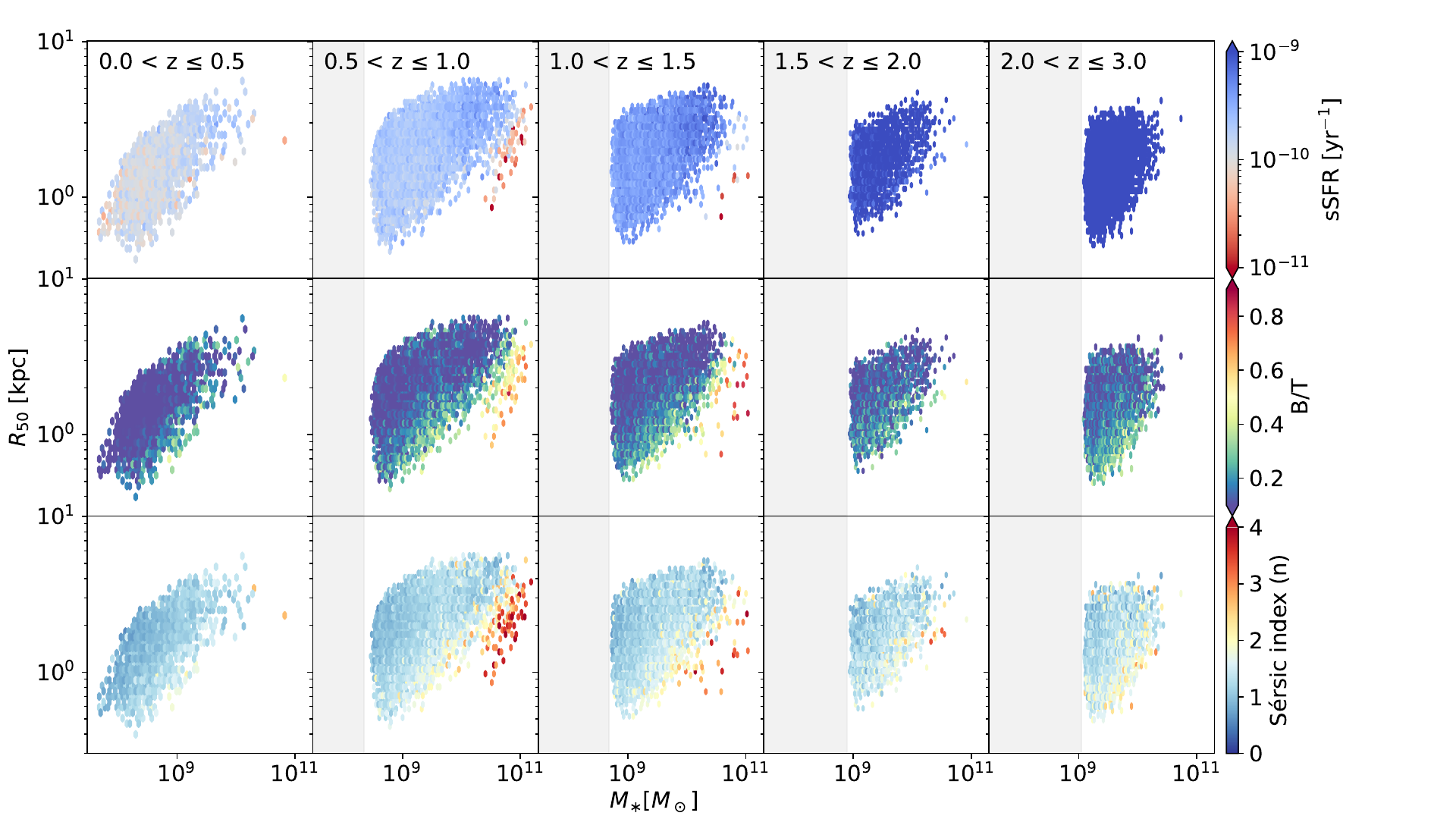} 
    \caption{Half-light radius $\hat{R}_{50}$~(physical size) vs. stellar mass $\hat{M}_{\ast}$ split into five redshift bins.  Cells are shaded by median sSFR \textit{(top)}, bulge-to-total ratio (from the F115W filter for $z < 2.0$ and from the F150W filter for $z \geq 2.0$, \textit{middle}) and S\'{e}rsic index $n$ \textit{(bottom)}. The cells are only shaded when they contain more than 7 galaxies. The grey shaded area shows the \texttt{pop-cosmos} mass completeness limit evaluated at the centre of each redshift bin.}
    \label{fig:ms-r50-z-bins-cw-full}
\end{figure*}

Further sources are removed by excluding anomalously large, low-stellar-mass objects -- $8.6 < \hat{m}_{\ast} < 10.5 $ and $ 0.8 < \log_{10}{(\hat{R}_{50}}/~\mathrm{kpc})$, due to unreliability of their size measurements. This unreliability arises from modelling artefacts, such as sources that are fainter than the depth of one or more of the four modelling bands or may have brighter companions nearby (see section 5.1 in \citealp{shuntov25}). This leaves us with a sample of $110\,273$ objects. Finally, we further limit the sample to the redshift interval $0.0 < \hat{z} \leq 3.0$, resulting in a sample of $99\,369$, which we analyse for the remainder of this work. Figure \ref{fig:workflow} shows a schematic of the workflow used to obtain this final catalogue.

\section{Methodology}
\label{sec:methods}

Galaxy sizes depend on both stellar mass and redshift, and how we measure that dependence is sensitive to the assumed functional form of the scaling relations (Section~\ref{subsec:relations}) and the classification scheme (Section~\ref{subsec:classification}).

\subsection{Functional forms of the size scaling relations}
\label{subsec:relations}
The form of the galaxy size--mass relation we adopt is
\begin{equation}
    R_{50} = R_{0} \, \left(\frac{M_{\ast}}{5 \times 10^{10}\, \mathrm{M}_{\odot}}\right)^{B},
    \label{eqn:linear-sm}
\end{equation}
where $R_{0}$ is the fiducial size scale for galaxies with $M_{\ast} = 5 \times 10^{10}\,\mathrm{M}_{\odot}$, and $B$ is the logarithmic slope. We adopt this form for the star-forming/disc-dominated/late-type galaxies, as well as the high-mass end of the quiescent galaxy population and the bulge-dominated/early-type galaxies \citep[following][]{shen03, vanderwel14, dimauro19}. 

Following \cite{shen03}, \citet{vanderwel14}, and \citet{nedkova21}, when we split the galaxy sample based on their star formation activity, we fit the quiescent population (and subsequently, the bulge-dominated and the early-type galaxies) with a double-power law. This has the form
\begin{equation}
    R_{50} = \gamma \left(\frac{M_{\ast}}{\mathrm{M}_{\odot}}\right)^{\alpha}\left(1+\frac{M_*}{10^{\delta} \,\mathrm{M}_{\odot} }\right)^{\beta-\alpha},
    \label{eqn:dbpl-sm}
\end{equation}
where $\alpha$ is the low-mass slope, $\beta$ is the high-mass slope, $\gamma$ gives the normalisation, and $\delta$ is the logarithm of the pivot mass where the function transitions between the two asymptotic power laws (see \citealp{nedkova21}). 

For the size--redshift relation, we assume a functional form
\begin{equation}
    R_{0} = B_z(1+z_{\mathrm{m}})^{-\beta_z},
    \label{eqn:sz}
\end{equation}
where $R_{0}$ is the fiducial scale obtained in Equation \ref{eqn:linear-sm} for different redshift bins, $z_{\mathrm{m}}$ is the mean redshift of each bin, $B_z$ and $\beta_z$ are the intercept and the slope of the fit.

\begin{figure*} 
    \centering
    \captionsetup{justification=centering}
    \includegraphics[width=1\linewidth]{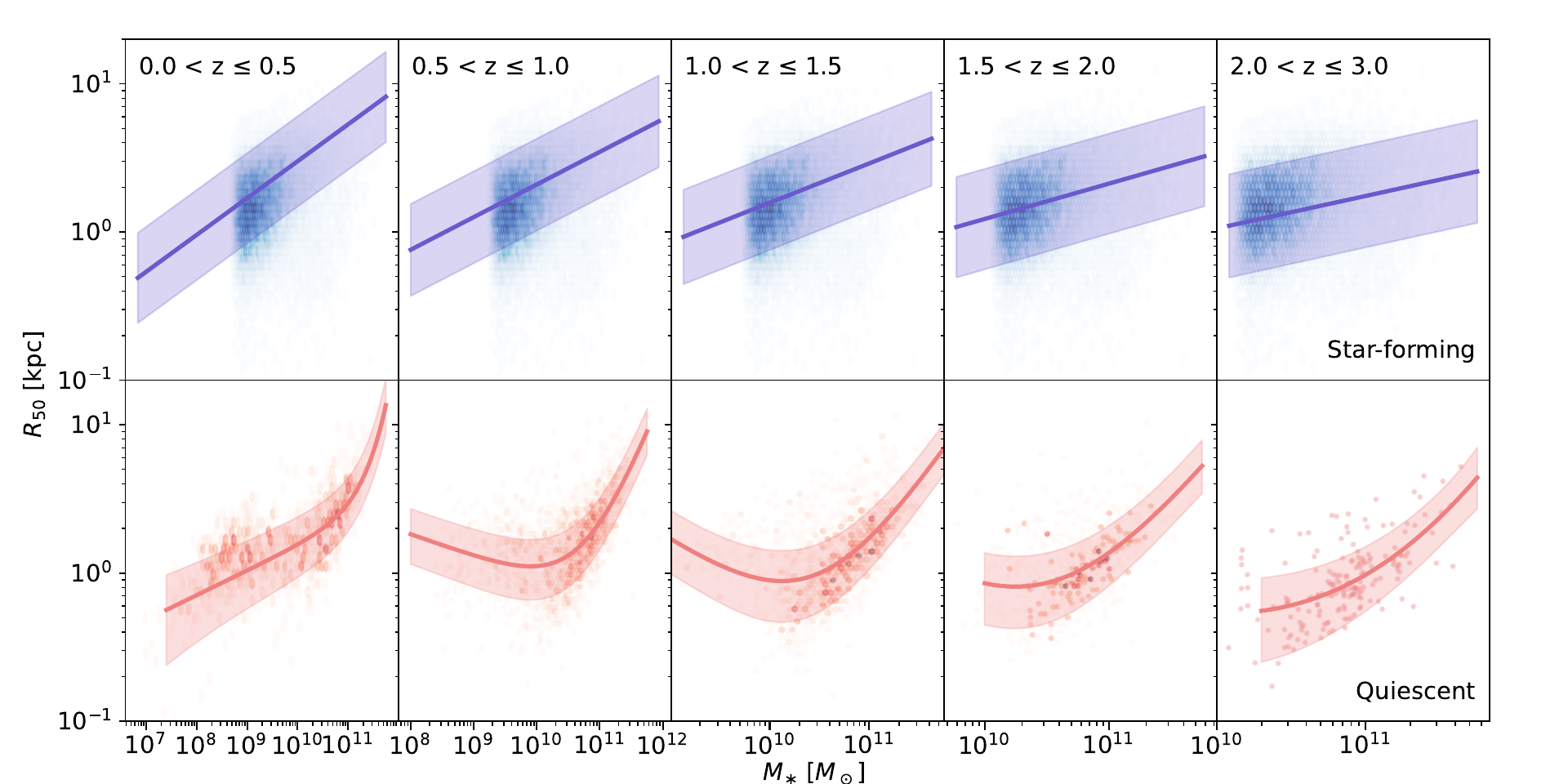} 
    \caption{Size--mass relation in five different redshift bins, split by star-formation activity. A single-power law is fit to the star-forming galaxies, defined as having $\log_{10}(\mathrm{sSFR}/~\mathrm{yr}^{-1}) > -11.0$ (\textit{top row}); a double-power law is fit to the quiescent galaxies with $\log_{10}(\mathrm{sSFR}/~\mathrm{yr}^{-1}) \leq -11.0$ (\textit{bottom row}). Cells are shaded by number of galaxies and are only shaded when they contain more than 3 galaxies. The shaded area shows the intrinsic scatter combined with the credible interval ($16^{\mathrm{th}}$ and $84^{\mathrm{th}}$ percentile) on the median fit. Due to the smaller number of objects, we show the last quiescent panel as a scatter plot of all galaxies instead of a hexbin, as in the other panels.}
    \label{fig:ms-r50-z-bins-sf-fit}
\end{figure*}

\subsection{Sample classifications and binning}
\label{subsec:classification}

To separate galaxies based on their star-formation activity we follow \citet{ilbert10, ilbert13} by defining: 
\begin{itemize}
    \item $\log_{10}{(\mathrm{sSFR}/\mathrm{yr}^{-1})} \leq -11$: quiescent galaxy;
    \item $\log_{10}{(\mathrm{sSFR}/\mathrm{yr}^{-1})} > -11$: star-forming galaxy.
\end{itemize}

We choose the sSFR criterion instead of the $UVJ$ or the $NUVrJ$ criteria because it provides a cleaner separation at low sSFR (see Section \ref{subsec:interchange}): the $UVJ$ selection cannot make distinctions below sSFR $< 10^{-10.5} \mathrm{yr}^{-1}$ \citep{leja19_uvj}; and the $NUVrJ$ selection suffers from $\sim20$~per cent contamination by dusty star-forming galaxies with sSFR $> 10^{-11} \mathrm{yr}^{-1}$ \citep{deger25}. We also use the same sSFR-split threshold in all redshift bins, since a redshift-dependent sSFR-split makes a minimal difference across the redshift range considered in this work, as discussed in \citet{deger25}. Additionally, we perform the single-power law fitting of the quiescent sample to a high mass range, such that $M_{\ast,q} > 10^{10.3}\,\mathrm{M}_{\odot}$, as other studies apply the same cut \citep[e.g.][]{vanderwel14, nedkova21}. For consistency, we do the same for the bulge-dominated and early-type galaxy samples.

For the bulge-to-total ratio (B/T), the galaxies are split such that \citep{yang25}:
\begin{itemize}
    \item B/T $\leq 0.2$: disc-dominated galaxy;
    \item $0.2 <$ B/T $< 0.6$: intermediate galaxy;
    \item  B/T $ \geq 0.6$: bulge-dominated galaxy.
\end{itemize}
Finally, as a function of S\'{e}rsic index $\hat{n}$, we split the galaxies into two samples \citep{lange15}:
\begin{itemize}
    \item $\hat{n} \leq 2.5$: late-type galaxy;
    \item $\hat{n} > 2.5$: early-type galaxy.
\end{itemize}

We now describe the details of the binning and splitting of the sample in order to fit Equations \ref{eqn:linear-sm}, \ref{eqn:dbpl-sm} and \ref{eqn:sz}. For the size--mass relation (Equations \ref{eqn:linear-sm} and \ref{eqn:dbpl-sm}), we split the sample in five redshift bins of width $\Delta z = 0.5$ from redshift $z=0$ to redshift $z=3$. For the size--redshift relation (Equation \ref{eqn:sz}), we use the same redshift binning, centred on the mean redshift of each bin. Within these bins, we ensure that the sample is mass complete and there are no selection effects introduced by the crossmatch, as discussed in Section \ref{sec:cross-match} and Appendix \ref{sec:completeness}.

\subsection{Inference of size scaling relations}
\label{subsec:fitting}

We assume a Gaussian likelihood for a galaxy of size $\hat{r}_{50} \equiv \log_{10}{(\hat{R}_{50}~/~\mathrm{kpc}})$ given a stellar mass $\hat{m}_{\ast} = \log_{10}{(\hat{M}_{\ast}~/~\mathrm{M}_{\odot})}$. Hence, 
\begin{equation}
    P\left(\hat{r}_{50} \mid \hat{m}_{\ast}\right)= \frac{1}{\sqrt{2 \pi}  \,\sigma} \exp \left[- \frac{1}{2} \, \frac{\left(\hat{r}_{50}-r_{50}\right)^2}{ \sigma^2}\right],
\end{equation} 
where the hat is used to indicate measured quantities, $r_{50}$ is the predicted half-light radius from either Equation~\ref{eqn:linear-sm} or~\ref{eqn:dbpl-sm}, and $\sigma$ is defined as (following \citealt{abdullah25}): 
\begin{equation}
    \sigma^2 = B^{2}\hat{\sigma}^{2}_{{\hat{m}_{\ast}}} + \hat{\sigma}^2_{\hat{r}_{50}} + \sigma^2_{\mathrm{int}}
\end{equation}
for the single-power law fit, and
\begin{equation}
    \sigma^2 = \bigg[\alpha+(\beta-\alpha)\frac{M_{\ast}}{10^{\delta}M_{\odot}+M_{\ast}}\bigg]^{2}\hat{\sigma}^{2}_{{\hat{m}_{\ast}}} + \hat{\sigma}^2_{\hat{r}_{50}} + \sigma^2_{\mathrm{int}}
\end{equation}
for the double-power law fit. Here, $\hat{\sigma}_{\hat{m}_{\ast}}$ is the uncertainty on the base–10 logarithm of the stellar mass, defined as the  posterior standard deviation from \texttt{pop-cosmos}, $\hat{\sigma}_{\hat{r}_{50}} \equiv (1/\hat{R}_{50})\sigma_{\hat{R}_{50}}$ is the uncertainty on the base–10 logarithm of the half-light radius provided by the COSMOS2025 catalogue, and $\sigma_{\mathrm{int}}$ is the intrinsic scatter, which we treat as a free parameter to be fit. The pre-factor in front of the stellar mass uncertainty comes from the derivative of the half-light radius $R_{50}$ from Equation \ref{eqn:dbpl-sm} with respect to the base–10 logarithm of stellar mass $m_{\ast}$.

\begin{figure*} 
    \centering
    \captionsetup{justification=centering}
    \includegraphics[width=1\linewidth]{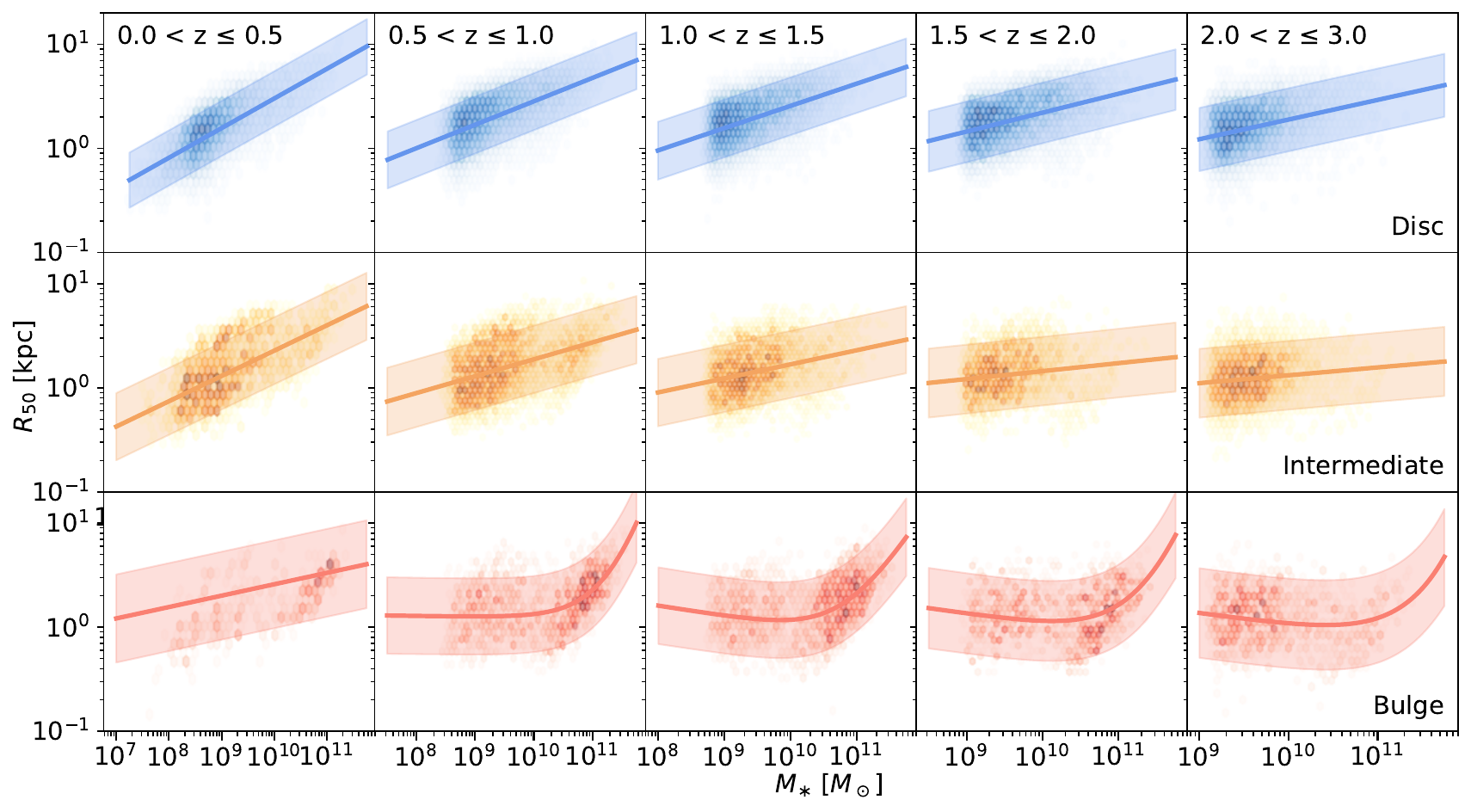} 
    \caption{Size--mass relation in five different redshift bins for the sample, separated by the bulge-to-total ratio computed from the F115W filter (for galaxies between $0.0 \leq z < 2.0)$ and the F150W filter (for galaxies between $2.0 \leq z \leq 3.0$). The sample is split into disc-dominated (B/T $ \leq 0.2$, \textit{top}), intermediate ($0.2 < $ B/T $<0.6$, \textit{middle}) and bulge-dominated ($0.6 \leq$ B/T, \textit{bottom}) galaxies. Cells are shaded by number of galaxies. A single-power law fit is performed on the disc-dominated and intermediate galaxies, whilst a double-power law is fitted to the bulge-dominated galaxies. The shaded area shows the intrinsic scatter combined with the credible interval ($16^{\mathrm{th}}$ and $84^{\mathrm{th}}$ percentile) on the median fit.}
    \label{fig:ms-r50-z-bins-bt-fit}
\end{figure*}

For the fitting of the size--redshift relation, we also use a Gaussian likelihood of the form
\begin{equation}
    P\left(\hat{R}_{0,z} \mid \hat{z}_m\right)= \frac{1}{\sqrt{2 \pi}  \,\sigma} \exp \left[- \frac{1}{2} \, \frac{\left(\hat{R}_{0,z}-R_{0,z}\right)^2}{ \sigma^2}\right],
\end{equation} 
where $\hat{R}_{0, z}$ is the fiducial scale in redshift bin $z_{\mathrm{m}}$ calculated from fitting Equation \ref{eqn:linear-sm}, $R_{0, z}$ is the fiducial scale given by Equation~\ref{eqn:sz} and the variance $\sigma$ is the width of the 68 per cent credible interval of $\hat{R}_{0, z}$ in the redshift bin $z_{\mathrm{m}}$. The effect of the intrinsic scatter from the size--mass relation is folded in by using the uncertainty of $\hat{R}_{0}$, since a large intrinsic scatter on the linear size--mass relation would lead to a less constrained $\hat{R}_0$, which yields less well-constrained size--redshift relations. 

Motivated by the results of previous studies such as \cite{lange15} and \cite{nedkova21}, we use uniform improper priors for the parameters, but with the scale radii ($R_0$ in Equation \ref{eqn:linear-sm} and $\gamma$ in Equation \ref{eqn:dbpl-sm}) constrained to be positive. We have checked that the data are sufficiently constraining to yield proper normalised posterior distributions.

We sample the posterior distributions of the parameters using \textsc{affine}\footnote{\url{https://github.com/justinalsing/affine/tree/torch}}, a GPU-accelerated affine-invariant Markov Chain Monte Carlo (MCMC) ensemble sampler, implemented in \textsc{PyTorch} \citep{paszke19} and based on the algorithm of \cite{goodman10}, with an improved implementation from \citet{foreman13}. We set $1000$ random walkers and let them run for $2500$ steps for all the single-power law fits, and $10000$ steps for all the double-power law fits, removing the first $100$ steps as burn-in. We compute the integrated autocorrelation time $\tau_f$ for each parameter, with values of $\sim35$--$42$ for the parameters of the linear size--mass relation, $\sim53$--$341$ for the parameters of the double-power law size--mass relation, and $\sim30$--$32$ for the parameters of the size--redshift relation\footnote{The largest values occur for the parameters that are prior-dominated in the least-constrained bins (the low-mass slope $\alpha$ and pivot mass $\delta$ where the break is unresolved; see Table \ref{tab:full-fit-dbpl}), rather than reflecting poor mixing of the well-constrained parameters. In every case, the chains are run for many autocorrelation times after burn-in: even the most autocorrelated parameter ($\tau_f \sim 341$) is sampled for $\sim 29$ autocorrelation times per walker.}. For all the parameters we summarise the constraints with the posterior median and the central 68 per cent credible interval ($16^{\mathrm{th}} - 84^{\mathrm{th}}$ percentile).

\section{Results}
\label{sec:results}

A key aim of this work is to measure the differences in size-scaling relations produced by sSFR-based (as opposed to colour–colour; Section \ref{subsec:interchange}) and morphology-based classifications, and investigate their causes. Here, we present the size--mass (Section \ref{subsec:size-mass}) and size--redshift (Section \ref{subsec:size-redshift}) relations under each classification, and assess robustness to Eddington and Malmquist bias (Section \ref{subsec:biases}).

\subsection{Size--mass relationship}
\label{subsec:size-mass}

First, we show the full size--mass relation in Figure \ref{fig:ms-r50-cw} for both the angular size (left) and the physical size (right). The results of the physical size fit are shown in Figure \ref{fig:ms-r50-z-bins-fit}, and the parameters presented in Table \ref{tab:full-fit}. We show the fits with solid lines, whilst the shaded regions show the intrinsic scatter $\sigma_{\mathrm{int}}$ combined with the posterior credible interval ($16^{\mathrm{th}}$ and $84^{\mathrm{th}}$ percentile). Both the fiducial scale $\log_{10}{(R_{0}~/~\mathrm{kpc})}$ and the slope $B$ decrease with redshift, albeit at different rates -- with the fiducial scale varying more strongly. The intrinsic scatter $\sigma_{\mathrm{int}}$ (in dex) is roughly constant across all redshift bins.

\begin{figure*} 
    \centering
    \captionsetup{justification=centering}
    \includegraphics[width=1\linewidth]{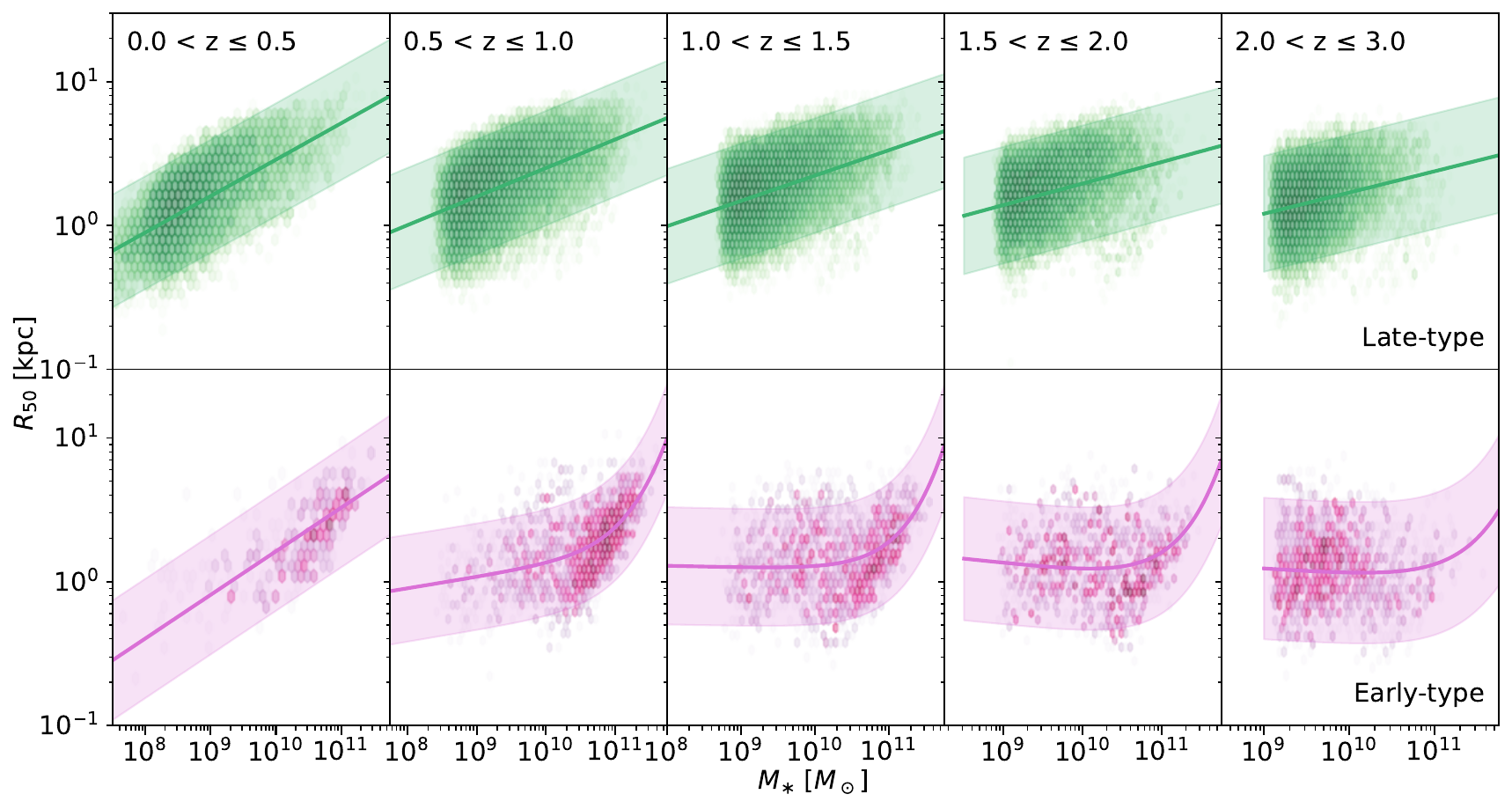} 
    \caption{Size--mass relation in five different redshift bins, split by S\'{e}rsic index. The sample is split into late- (\textit{top row}) and early-type (\textit{bottom row}) galaxies at S\'{e}rsic index $n = 2.5$. A single-power law fit is performed for the late-type sample, whilst a double-power law fit is performed for the early-type galaxies -- with the exception of the lowest redshift bin, where we show the single-power law fit. The shaded area shows the intrinsic scatter combined with the credible interval ($16^{\mathrm{th}}$ and $84^{\mathrm{th}}$ percentile) on the median fit.}
    \label{fig:ms-r50-z-bins-n-fit}
\end{figure*}

In the rest of this section, we split the sample by sSFR, B/T and S\'{e}rsic index (as defined in Section \ref{subsec:classification}), as shown in the three panels of Figure~\ref{fig:ms-r50-z-bins-cw-full}. Results for the sSFR, B/T, and S\'{e}rsic index splits are shown in Figures~\ref{fig:ms-r50-z-bins-sf-fit}, \ref{fig:ms-r50-z-bins-bt-fit}, and \ref{fig:ms-r50-z-bins-n-fit} respectively. The corresponding fitted parameters are given in Tables~\ref{tab:full-fit}, \ref{tab:bt_cut_fit}, \ref{tab:n-cut-fit} and \ref{tab:full-fit-sf} for the single-power law fits, and in Table~\ref{tab:full-fit-dbpl} for the double-power law fits. Single-power law fits to the high-mass quiescent populations are presented for comparison with previous work, as motivated in Section \ref{subsec:classification}. We also present single-power law fits to the high-mass ($M_\ast > 10^{10.3}\,\mathrm{M_\odot}$) bulge-dominated and early-type samples to enable direct comparison of the high-mass slope across all classification schemes on the same functional form. Across all three classifications, the size--mass relation exhibits the same broad, redshift-dependent behaviour: the intercept (fiducial scale, $\log_{10}{(R_0~/~\mathrm{kpc})}$) and the slope ($B$) both decline with increasing redshift, and this evolution is strongest for the fiducial scale. The intrinsic scatter, $\sigma_{\mathrm{int}}$, is approximately constant with redshift in every split and is of order $0.12-0.25$~dex. 

\begin{table*} 
    \centering
    \caption{Posterior median and 68 per cent credible interval ($16^\mathrm{th} - 84^{\mathrm{th}}$ percentile) for the parameters describing the linear size--mass relation for the whole sample (\textit{top}), the high-mass quiescent sample (\textit{middle}), and the star-forming population (\textit{bottom}) for the COSMOS-Web sample across five redshift bins.}
    \label{tab:full-fit}
    \begin{tabular}{lccccc}
    \toprule 
    sample & redshift bin & $N_{\mathrm{gal}}$ & \multicolumn{1}{c|}{$\log_{10}{(R_0/\mathrm{kpc}})$} & \multicolumn{1}{c}{$B$} & \multicolumn{1}{c|}{$\sigma_{\mathrm{int}}$~/dex}\\ \midrule 
    full sample& $0.0 \leq z < 0.5$ & $11364$ & \multicolumn{1}{c|}{$0.59 \pm 0.01$} & \multicolumn{1}{l|}{$0.23 \pm 0.01$} & \multicolumn{1}{c|}{$0.16 \pm 0.01$} \\ 
    & $0.5 \leq z < 1.0$ & $27666$ & \multicolumn{1}{c|}{$0.47 \pm 0.01$} & \multicolumn{1}{l|}{$0.16 \pm 0.01$} & \multicolumn{1}{c|}{$0.16 \pm 0.01$} \\ 
    & $1.0 \leq z < 1.5$ & $24830$ & \multicolumn{1}{c|}{$0.40 \pm 0.01$} & \multicolumn{1}{l|}{$0.13 \pm 0.01$} & \multicolumn{1}{c|}{$0.17 \pm 0.01$} \\ 
    & $1.5 \leq z < 2.0$ & $16197$ &\multicolumn{1}{c|}{$0.32 \pm 0.01$} & \multicolumn{1}{l|}{$0.10 \pm 0.01$} & \multicolumn{1}{c|}{$0.17 \pm 0.01$} \\
    & $2.0 \leq z \leq 3.0$ & $19312$ & \multicolumn{1}{c|}{$0.26 \pm 0.01$} & \multicolumn{1}{l|}{$0.10 \pm 0.01$} & \multicolumn{1}{c|}{$0.17 \pm 0.01$} \\ \midrule
    quiescent: $\log_{10}(\mathrm{sSFR}/\mathrm{yr}^{-1}) \leq -11$ \& $m_{\ast} > 10.3$ & $0.0 \leq z < 0.5$ & $304$ & \multicolumn{1}{c|}{$0.38 \pm 0.02$} & \multicolumn{1}{l|}{$0.56 \pm 0.07$} & \multicolumn{1}{c|}{$0.12 \pm 0.01$} \\ 
    & $0.5 \leq z < 1.0$ & $1315$ & \multicolumn{1}{c|}{$0.20 \pm 0.01$} & \multicolumn{1}{c|}{$0.56 \pm 0.03$} & \multicolumn{1}{c|}{$0.13 \pm 0.01$} \\ 
    & $1.0 \leq z < 1.5$ & $898$ & \multicolumn{1}{c|}{$0.09 \pm 0.01$} & \multicolumn{1}{c|}{$0.55 \pm 0.05$} & \multicolumn{1}{c|}{$0.15 \pm 0.01$} \\ 
    & $1.5 \leq z < 2.0$ & $529$ & \multicolumn{1}{c|}{$0.01 \pm 0.02$} & \multicolumn{1}{c|}{$0.52 \pm 0.06$} & \multicolumn{1}{c|}{$0.15 \pm 0.01$} \\
    & $2.0 \leq z \leq 3.0$ & $166$ & \multicolumn{1}{c|}{$-0.17 \pm 0.04$} & \multicolumn{1}{c|}{$0.62 \pm 0.11$} & \multicolumn{1}{c|}{$0.15 \pm 0.02$} \\ \midrule
    star-forming: $\log_{10}(\mathrm{sSFR}/\mathrm{yr}^{-1}) > -11$ & $0.0 \leq z < 0.5$ & $10427$ & \multicolumn{1}{c|}{$0.65 \pm 0.01$} & \multicolumn{1}{c}{$0.26 \pm 0.01$} & \multicolumn{1}{c}{$0.15 \pm 0.01$}\\ 
    & $0.5 \leq z < 1.0$ & $25686$ & \multicolumn{1}{c|}{$0.54 \pm 0.01$} & \multicolumn{1}{c}{$0.20 \pm 0.01$} & \multicolumn{1}{c}{$0.15 \pm 0.01$} \\ 
    & $1.0 \leq z < 1.5$ & $23711$ & \multicolumn{1}{c|}{$0.46 \pm 0.01$} & \multicolumn{1}{c}{$0.17 \pm 0.01$} & \multicolumn{1}{c}{$0.16 \pm 0.01$} \\ 
    & $1.5 \leq z < 2.0$ & $15616$ & \multicolumn{1}{c|}{$0.37 \pm 0.01$} &  \multicolumn{1}{c}{$0.13 \pm 0.01$}&  \multicolumn{1}{c}{$0.17 \pm 0.01$}\\
    & $2.0 \leq z \leq 3.0$ & $19135$ & \multicolumn{1}{c|}{$0.29 \pm 0.01$} &  \multicolumn{1}{c}{$0.11 \pm 0.01$}&  \multicolumn{1}{c}{$0.17 \pm 0.01$}\\
    \bottomrule
    \end{tabular}
\end{table*}

\begin{table*} 
    \centering
    \caption{Posterior median and 68 per cent credible interval ($16^\mathrm{th} - 84^{\mathrm{th}}$ percentile) for the parameters of the linear size--mass relation for the three bulge-to-total ratio ($\mathrm{B/T}$) populations for the COSMOS-Web sample across five redshift bins.}
\label{tab:bt_cut_fit}
\begin{tabular}{lccccc}
\toprule 
 sample & redshift bin & $N_{\mathrm{gal}}$ & \multicolumn{1}{c|}{$\log_{10}({R_0}/{\mathrm{kpc}})$} & \multicolumn{1}{c|}{$B$} & \multicolumn{1}{c|}{$\sigma_{\mathrm{int}}$~/dex} \\ \midrule
bulge-dominated: $\mathrm{B/T}\geq0.6$ \& $m_{\ast} > 10.3$ & $0.0 \leq z < 0.5$ & $205$ &$0.33 \pm 0.01$ & $0.57 \pm 0.03$ & $0.18 \pm 0.01$  \\ 
& $0.5 \leq z < 1.0$ & $1099$ & $0.22 \pm 0.01$ & $0.58 \pm 0.02$ & $0.16 \pm 0.01$  \\ 
& $1.0 \leq z < 1.5$ & $1519$ &$0.19 \pm 0.01$ & $0.52 \pm 0.02$  & $0.16 \pm 0.01$  \\ 
& $1.5 \leq z < 2.0$ & $1016$ & $0.09 \pm 0.01$ & $0.53 \pm 0.03$  & $0.16 \pm 0.01$   \\ 
& $2.0 \leq z \leq 3.0$ & $545$ & $0.01 \pm 0.01$ & $0.37 \pm 0.04$ & $0.20 \pm 0.01$ \\ \midrule
intermediate: $0.2 < \mathrm{B/T} < 0.6$ & $0.0 \leq z < 0.5$ & $2449$ &$0.53 \pm 0.02$ & $0.24 \pm 0.01$ & $0.17 \pm 0.01$ \\
& $0.5 \leq z < 1.0$ & $6528$ & $0.39 \pm 0.01$ & $0.16 \pm 0.01$ & $0.18 \pm 0.01$ \\
& $1.0 \leq z < 1.5$ & $5358$ & $0.32 \pm 0.01$ & $0.13 \pm 0.01$ & $0.17 \pm 0.01$ \\
& $1.5 \leq z < 2.0$ & $3590$ &$0.21 \pm 0.02$ & $0.07 \pm 0.02$ & $0.18 \pm 0.01$ \\
& $2.0 \leq z \leq 3.0$ & $4863$ & $0.17 \pm 0.02$ & $0.07 \pm 0.02$ & $0.18 \pm 0.01$\\ \midrule
disc-dominated: $\mathrm{B/T}\leq0.2$ & $0.0 \leq z < 0.5$ & $8256$ & $0.70 \pm 0.01$ & $0.27 \pm 0.01$ & $0.13 \pm 0.01$ \\
& $0.5 \leq z < 1.0$ & $18741$ & $0.61 \pm 0.01$ & $0.22 \pm 0.01$ & $0.14 \pm 0.01$ \\
& $1.0 \leq z < 1.5$ & $16574$& $0.56 \pm 0.01$ & $0.21 \pm 0.01$ & $0.14 \pm 0.01$ \\
& $1.5 \leq z < 2.0$ & $10577$ &$0.47 \pm 0.01$ & $0.18 \pm 0.01$ & $0.14 \pm 0.01$ \\
& $2.0 \leq z \leq 3.0$ & $12556$& $0.41 \pm 0.01$ & $0.19 \pm 0.01$ & $0.15 \pm 0.01$ \\
\hline
\end{tabular}
\end{table*}

However, the three splits show differences in several important ways. First, the fitted slopes differ systematically between classes: star-forming/disc-dominated/late-type samples have shallower slopes (typical $B \sim 0.11$--$0.27$ depending on redshift -- Tables \ref{tab:full-fit}--\ref{tab:n-cut-fit}), whereas high-mass quiescent/bulge-dominated/early-type samples show on average a constant, steeper high-mass behaviour -- for example, the quiescent slope centres near $B \sim 0.55$ in the linear high-mass fits (Table~\ref{tab:full-fit}), $B \sim 0.54$ for the bulge-dominated fits (Table~\ref{tab:bt_cut_fit}), or $B \sim 0.46$ for the early-type fits (Table~\ref{tab:n-cut-fit}). The linear scale radii $R_0$ also differ between the two classes, with the star-forming/disc-dominated/late-type systems having a higher scale radius than the quiescent/bulge-dominated/early-type galaxies, irrespective of the stellar mass range (Tables \ref{tab:full-fit}--\ref{tab:full-fit-sf}).

Second, the intrinsic scatter depends on the structural split: bulge-dominated and intermediate B/T populations exhibit larger intrinsic scatter $\sigma_{\mathrm{int}}$ than disc-dominated systems ($\sigma_{\mathrm{int}} = 0.13$--$0.15$~dex for disc-dominated systems vs. $\sigma_{\mathrm{int}} = 0.19$--$0.22$~dex for bulge-dominated -- Tables \ref{tab:bt_cut_fit}, \ref{tab:full-fit-sf} and \ref{tab:full-fit-dbpl}), and similarly, early-type galaxies show higher intrinsic scatter than late-types ($\sigma_{\mathrm{int}} = 0.14$--$0.16$~dex for late-type systems vs. $\sigma_{\mathrm{int}} = 0.16$--$0.22$~dex for early-type -- Tables \ref{tab:n-cut-fit} and \ref{tab:full-fit-dbpl}). By contrast, when splitting by sSFR the intrinsic scatter $\sigma_{\mathrm{int}}$ for star-forming and quiescent populations is broadly similar and approximately constant with redshift ($\sigma_{\mathrm{int}} \sim 0.15$~dex -- Tables \ref{tab:full-fit} and \ref{tab:full-fit-dbpl}). 

Third, the low-mass behaviour of the relations differs. When quiescent (or bulge-dominated/early-type) systems are fitted with Equation \ref{eqn:dbpl-sm}, the low-mass slope is substantially shallower than the high-mass slope (e.g. $\alpha = -0.45$--$0.15$ compared to $\beta = 0.94$--$1.07$ for the quiescent fits, where constrained) and the pivot mass ($\delta$) that indicates the transition varies between fits; this fracture is less apparent for the disc-dominated/late-type samples, which are well described by a single-power law across the mass range. The double-power law fits reveal that the pivot masses $\delta$ are only well-constrained over a restricted redshift range. At $0.5 \leq z < 1.5$, both the quiescent and bulge-dominated samples yield well-located pivots that are stable across the two bins (differing by $1.0~\sigma$ and $1.6~\sigma$, respectively), giving $\delta_{\rm Q} \sim 10.7 \pm 0.2$~dex and $\delta_{\rm B} \sim 11.1 \pm 0.2$~dex over this window. At $z \geq 1.5$, the quiescent fits are unconstrained, and the bulge-dominated fits return lower limits consistent with no required break below $M_{\ast} \sim 10^{11.4}~\mathrm{M}_{\odot}$; in both cases the data are insufficient to locate a transition. At $z < 0.5$, the fits yield lower limits that are inconsistent with the mid-redshift pivot locations: for the quiescent sample $\delta_{\rm Q} > 11.6$~dex, which lies more than $0.9$~dex above the $0.5 \leq z < 1.5$ value, and for the bulge-dominated sample $\delta_{\rm B} > 11.7$~dex, more than $0.3$~dex above the adjacent bin. This suggests that below $z = 0.5$ the relation either ceases to be well described by a broken power law, or the transition has migrated to masses beyond the range sampled here; we return to this in Section \ref{subsubsec:agn}. The early-type ($n > 2.5$) sample yields no located pivot in any redshift bin: the lower limits on $\delta$ are consistent with the bulge-dominated values at $0.5 \leq z < 1.0$ but lie above them at $1.0 \leq z < 1.5$, and the data are consistent with a single-power law throughout. We discuss this finding in Section \ref{subsec:interchange} and Appendix \ref{bt-vs-n}. 

The scale radii reported in Tables \ref{tab:full-fit-sf} and \ref{tab:full-fit-dbpl} provide a consistency check between the linear and double-power law cases. In Table \ref{tab:full-fit-dbpl}, we report $\log_{10}{(R_{\ast}/\mathrm{kpc})}$ evaluated at $M_{\ast} = 5 \times 10^{10}~M_{\odot}$ rather than the normalisation $\gamma$ (see Equation \ref{eqn:dbpl-sm}) directly, to facilitate comparison with the linear fiducial scale $\log_{10}{(R_{0}/\mathrm{kpc})}$ in Table \ref{tab:full-fit-sf}. The reference mass $5 \times 10^{10}~\mathrm{M}_{\odot}$ ($m_{\ast} = 10.70$) sits around the quiescent pivot masses ($\delta_\mathrm{Q} \sim 10.5$--$10.8$~dex), so this column reports the size approximately at the pivot mass -- the point where the double-power law transitions. At $z \geq 1.5$, where $\delta$ is unconstrained and $\alpha$ is undetermined, the double-power law normalisation $R_{\ast}$ is still recovered (Table \ref{tab:full-fit-dbpl}), since evaluating at the reference mass marginalises over the $\alpha-\delta$ degeneracy \citep[see][]{nedkova21}: the data pin the size at $5 \times 10^{10}~\mathrm{M}_{\odot}$ wherever that mass falls within the observed range, regardless of whether the break is located. Comparing the two cases directly, the double-power law normalisation sits $0.04$--$0.07$~dex below the linear fiducial scale at $0.5 \leq z < 1.5$, where the break is constrained, at a significance of $6$--$8\sigma$ per bin. This offset has the expected sign: at $5 \times 10^{10}~\mathrm{M}_{\odot}$, which lies at or above the pivot, the relation has bent onto its steep high-mass arm, hence the double-power law predicts a smaller size at this mass than the single-power law fit to the same subsample. The systematic offset in the two bins where the break is detected is therefore independent evidence that the break is real in those bins.

Finally, practical measurement and selection effects modulate these differences. The B/T measurement depends on band and resolution: by using a space-based B/T, we recover sharper trends with mass, redshift and sSFR than ground-based B/T. We will return to this point in Section \ref{subsec:interchange}.

\begin{table*} 
    \centering
    \caption{Posterior median and 68 per cent credible interval ($16^\mathrm{th} - 84^{\mathrm{th}}$ percentile) for the linear size--mass relation for the two S\'{e}rsic populations for the COSMOS-Web sample across five redshift bins.}
    \label{tab:n-cut-fit}
    \begin{tabular}{lccccc}
    \toprule 
     sample & redshift bin & $N_{\mathrm{gal}}$ & \multicolumn{1}{c|}{$\log_{10}{(R_0~/~\mathrm{kpc})}$} & \multicolumn{1}{c|}{$B$} & \multicolumn{1}{c|}{$\sigma_{\mathrm{int}}$~/dex}   \\ \midrule 
    late-type: $n \leq 2.5$ & $0.0 \leq z < 0.5$ & $10618$ & $0.64 \pm 0.02$ & $0.25 \pm 0.01$ & $0.14 \pm 0.01$  \\ 
    &$0.5 \leq z < 1.0$ & $24817$ & $0.54 \pm 0.01$ & $0.20 \pm 0.01$ & $0.15 \pm 0.01$  \\ 
    &$1.0 \leq z < 1.5$ & $22090$ & $0.47 \pm 0.01$ & $0.18 \pm 0.01$  & $0.15 \pm 0.01$ \\ 
    &$1.5 \leq z < 2.0$ & $13969$ & $0.40 \pm 0.01$ & $0.15 \pm 0.01$  & $0.16 \pm 0.01$ \\
    &$2.0 \leq z \leq 3.0$ & $16728$ & $0.33 \pm 0.01$ & $0.15 \pm 0.01$  & $0.16 \pm 0.01$ \\ \midrule
    early-type: $n > 2.5$ \& $m_{\ast} > 10.3$ & $0.0 \leq z < 0.5$ & $350$ & $0.39 \pm 0.02$ & $0.55 \pm 0.06$ & $0.16 \pm 0.01$ \\ 
    &$0.5 \leq z < 1.0$ & $1660$ & $0.26 \pm 0.01$ & $0.50 \pm 0.03$ & $0.18 \pm 0.01$  \\ 
    &$1.0 \leq z < 1.5$ & $1290$ & $0.17 \pm 0.01$ & $0.46 \pm 0.05$  & $0.17 \pm 0.01$ \\ 
    &$1.5 \leq z < 2.0$ & $981$ & $0.12 \pm 0.02$ & $0.39 \pm 0.06$  & $0.19 \pm 0.01$  \\
    &$2.0 \leq z \leq 3.0$ & $636$ & $0.07 \pm 0.02$ & $0.31 \pm 0.09$  & $0.22 \pm 0.01$  \\
    \bottomrule
    \end{tabular}
\end{table*}

\subsection{Size--redshift relationship}
\label{subsec:size-redshift}

\begin{figure*} 
    \centering
    \captionsetup{justification=centering}
    \includegraphics[width=1\linewidth]{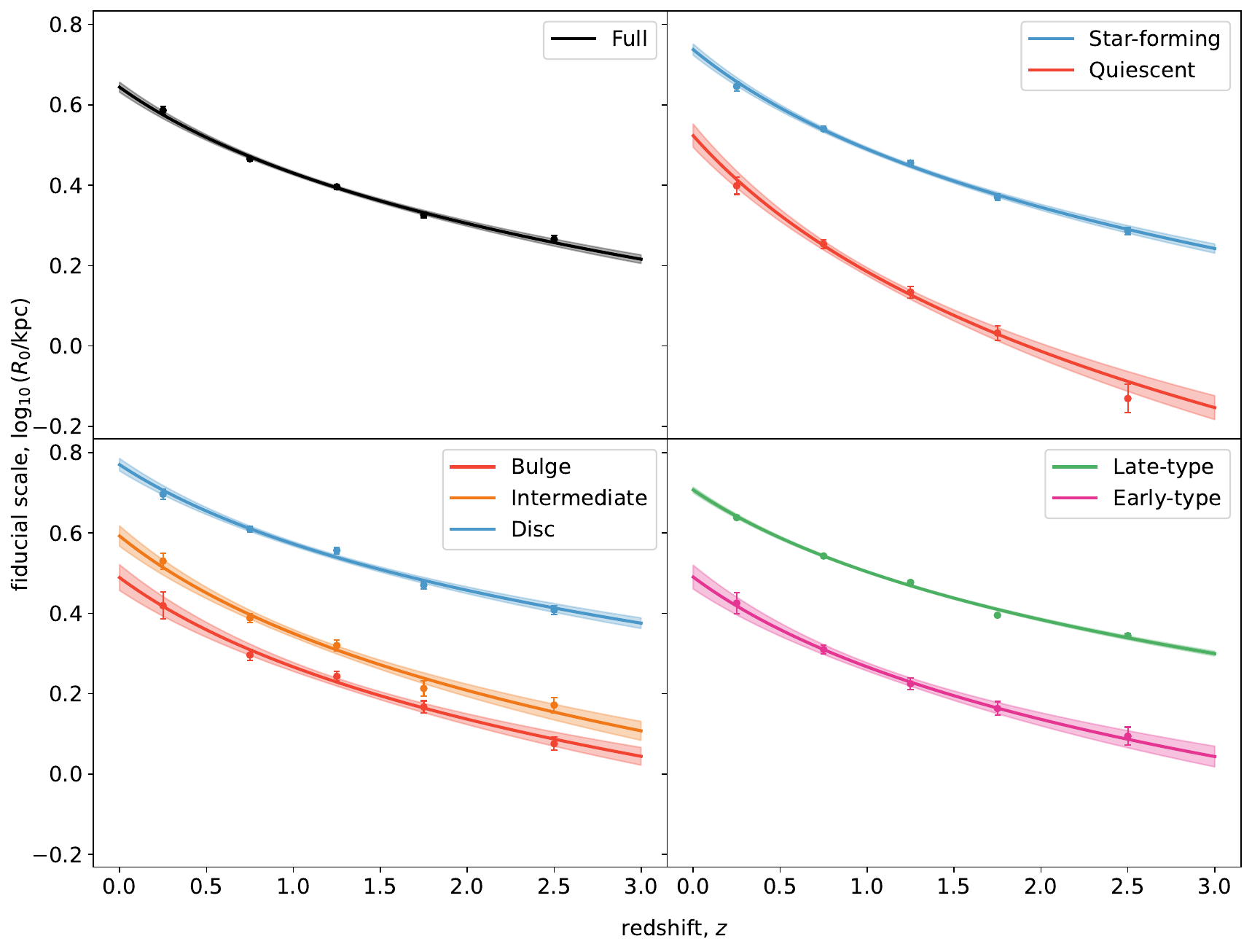} 
    \caption{The size--redshift relation for the four different cases: full sample (\textit{top left}), star-forming/quiescent split (\textit{top right}), B/T split (\textit{bottom left}) and S\'{e}rsic index split (\textit{bottom right}). For the star-forming/quiescent classification, the sample is split at $\log_{10}{(\mathrm{sSFR}/\mathrm{yr}^{-1})} \leq -11$. For the bulge-to-total ratio classification, the sample is split into disc-dominated (B/T $\leq 0.2$), intermediate ($0.2 < $ B/T $<0.6$) and bulge-dominated ($0.6 \leq$ B/T) galaxies. For the S\'{e}rsic index classification, the sample is split into early and late galaxies at $n = 2.5$. The solid lines show the fits to Equation \ref{eqn:sz}, whilst the shaded areas show the 68 per cent credible interval ($16^{\mathrm{th}}$ and $84^{\mathrm{th}}$ percentile) on the median fit.}
    \label{fig:z-r50-fit}
\end{figure*}

Figure \ref{fig:z-r50-fit} and Table \ref{tab:size_redshift_fits} show the size--redshift relation for the full sample and for each of the three classification splits. In the figure, the solid line represents the posterior median of the fit, whilst the shaded area encloses the $16^{\mathrm{th}}$ and $84^{\mathrm{th}}$ percentiles of the posterior. 

\begin{table*} 
    \centering
    \caption{Posterior median and 68 per cent credible interval ($16^\mathrm{th} - 84^{\mathrm{th}}$ percentile) for the parameters describing the linear size--mass relation for the whole quiescent sample (\textit{top}), bulge-dominated sample (\textit{middle}), and the early-type population (\textit{bottom}) for the COSMOS-Web sample across five redshift bins.}
    \label{tab:full-fit-sf}
    \begin{tabular}{lccccc}
    \toprule 
    sample & redshift bin & $N_{\mathrm{gal}}$ & \multicolumn{1}{c|}{$\log_{10}{(R_0/\mathrm{kpc}})$} & \multicolumn{1}{c}{$B$} & \multicolumn{1}{c|}{$\sigma_{\mathrm{int}}$~/dex}\\ \midrule 
    quiescent: $\log_{10}(\mathrm{sSFR}/\mathrm{yr}^{-1}) \leq -11$ & $0.0 \leq z < 0.5$ & $937$ & \multicolumn{1}{c|}{$0.40 \pm 0.02$} & \multicolumn{1}{l|}{$0.20 \pm 0.01$} & \multicolumn{1}{c|}{$0.16 \pm 0.01$} \\ 
    & $0.5 \leq z < 1.0$ & $1980$ & \multicolumn{1}{c|}{$0.25 \pm 0.01$} & \multicolumn{1}{l|}{$0.19 \pm 0.01$} & \multicolumn{1}{c|}{$0.16 \pm 0.01$} \\ 
    & $1.0 \leq z < 1.5$ & $1119$ & \multicolumn{1}{c|}{$0.13 \pm 0.01$} & \multicolumn{1}{l|}{$0.28 \pm 0.02$} & \multicolumn{1}{c|}{$0.17 \pm 0.01$} \\ 
    & $1.5 \leq z < 2.0$ & $581$ &\multicolumn{1}{c|}{$0.03 \pm 0.01$} & \multicolumn{1}{l|}{$0.39 \pm 0.03$} & \multicolumn{1}{c|}{$0.15 \pm 0.01$} \\
    & $2.0 \le z \leq 3.0$ & $177$ & \multicolumn{1}{c|}{$-0.13 \pm 0.02$} & \multicolumn{1}{l|}{$0.48 \pm 0.05$} & \multicolumn{1}{c|}{$0.15 \pm 0.01$} \\ \midrule
     bulge-dominated: B/T $\geq 0.6$ & $0.0 \leq z < 0.5$ & $659$ & \multicolumn{1}{c|}{$0.42 \pm 0.01$} & \multicolumn{1}{l|}{$0.24 \pm 0.06$} & \multicolumn{1}{c|}{$0.21 \pm 0.01$} \\ 
    & $0.5 \leq z < 1.0$ & $2397$ & \multicolumn{1}{c|}{$0.30 \pm 0.01$} & \multicolumn{1}{c|}{$0.14 \pm 0.01$} & \multicolumn{1}{c|}{$0.19 \pm 0.01$} \\ 
    & $1.0 \leq z < 1.5$ & $2898$ & \multicolumn{1}{c|}{$0.24 \pm 0.01$} & \multicolumn{1}{c|}{$0.13 \pm 0.02$} & \multicolumn{1}{c|}{$0.19 \pm 0.01$} \\ 
    & $1.5 \leq z < 2.0$ & $2030$ & \multicolumn{1}{c|}{$0.17 \pm 0.02$} & \multicolumn{1}{c|}{$0.08 \pm 0.02$} & \multicolumn{1}{c|}{$0.20 \pm 0.01$} \\
    & $2.0 \leq z \leq 3.0$ & $1893$ & \multicolumn{1}{c|}{$0.08 \pm 0.02$} & \multicolumn{1}{c|}{$0.01 \pm 0.01$} & \multicolumn{1}{c|}{$0.22 \pm 0.01$} \\ \midrule
    early-type: $n > 2.5$ & $0.0 \leq z < 0.5$ & $746$ & \multicolumn{1}{c|}{$0.43 \pm 0.03$} & \multicolumn{1}{c}{$0.30 \pm 0.02$} & \multicolumn{1}{c}{$0.21 \pm 0.01$}\\ 
    & $0.5 \leq z < 1.0$ & $2849$ & \multicolumn{1}{c|}{$0.31 \pm 0.01$} & \multicolumn{1}{c}{$0.20 \pm 0.01$} & \multicolumn{1}{c}{$0.19 \pm 0.01$} \\ 
    & $1.0 \leq z < 1.5$ & $2740$ & \multicolumn{1}{c|}{$0.22 \pm 0.01$} & \multicolumn{1}{c}{$0.11 \pm 0.02$} & \multicolumn{1}{c}{$0.21 \pm 0.01$} \\ 
    & $1.5 \leq z < 2.0$ & $2228$ & \multicolumn{1}{c|}{$0.16 \pm 0.02$} &  \multicolumn{1}{c}{$0.06 \pm 0.02$}&  \multicolumn{1}{c}{$0.22 \pm 0.01$}\\
    & $2.0 \leq z \leq 3.0$ & $2584$ & \multicolumn{1}{c|}{$0.09 \pm 0.02$} &  \multicolumn{1}{c}{$0.02 \pm 0.02$}&  \multicolumn{1}{c}{$0.25 \pm 0.01$}\\
    \bottomrule
    \end{tabular}
\end{table*}

Across every sample split (sSFR, B/T and S\'{e}rsic index), the qualitative behaviour is the same: galaxy sizes decrease with redshift, and this evolution is well described by Equation \ref{eqn:sz}. Star-forming (or late-type/disc-dominated) populations show systematically shallower evolution (smaller $\beta_z$) and larger normalisations $B_z$ than quenched (or early-type) populations, which evolve more rapidly in size (larger $\beta_z$). Intermediate B/T systems show relatively high $\beta_z$ values that exceed those of disc-dominated systems, while disc-dominated galaxies retain the highest $B_z$ (i.e.\ the largest median sizes at $z \approx 0$). The S\'{e}rsic index split mirrors this pattern: late-type systems have the shallowest evolution, and early-type systems evolve more steeply. 

When using the scale radius from the double-power law fits, $R_{50}(M_{\ast}=5 \times 10^{10}~\mathrm{M}_{\odot})$ (Table \ref{tab:full-fit-dbpl}), instead of the linear fiducial scale $R_0$ (Tables \ref{tab:full-fit}--\ref{tab:full-fit-sf}), the values of $B_z$ are consistent with each other. This is expected -- for example, across both parametrisations, $\log_{10}{(R_{\ast}/\mathrm{kpc})}$ declines monotonically from $\sim 0.4$ to $\sim -0.2$ over the full redshift interval for the quiescent sample, reflecting size growth at fixed stellar mass. The rate of this evolution, $\mathrm{d}\log_{10}{R}/\mathrm{d}z$, is $-0.23$ from the single-power law fit and $-0.24$ from the double-power law, consistent within errors, confirming that the rate of size growth at fixed mass is robust to the choice of functional form even though the zero-point is not. Further, both the bulge-dominated and the early-type cases follow the same pattern as the quiescent galaxies.

\subsection{Effects of various biases}
\label{subsec:biases}
Using the size--mass relation fits, we verify robustness to two potential biases: Eddington bias from both sSFR and stellar mass uncertainties; and Malmquist bias near the completeness limit.

For the sSFR-based split, Eddington bias could in principle scatter galaxies from the smaller quiescent population into the larger star-forming one, contaminating both sub-samples. We quantify this in Appendix~\ref{sec:ssfr_reliability} by computing the posterior probability that each galaxy's star-forming or quiescent assignment is correct. The effect is negligible: 92 per cent of star-forming galaxies are assigned to the star-forming class with greater than 95 per cent posterior probability, and the quiescent sample contains only a small number of low-confidence assignments.

For the Eddington bias from the stellar mass uncertainties, we compute an analytical correction for the slope $B$ and compare it to the uncertainties on the fit. Specifically, we calculate the log-slope of the stellar mass function, such that
\begin{equation}
    \epsilon(M_{\ast}) = (N_{i+1}-N_i)/\Delta \log_{10}{(M_{\ast}/\mathrm{M}_{\odot})},
\end{equation}
where $N_i$ is the number of galaxies in bin $i$ and $\Delta \log_{10}{(M_{\ast}/\mathrm{M_{\odot}})}$ is the size of the mass bin. We then compute a fractional bias estimate $\Delta B$ using the slope $B$, the slope computed from the mass function $\epsilon$ and the uncertainty in the stellar mass given as $\sigma_{m_{\ast}} = 0.12$, such that
\begin{equation}
    \Delta B = B \times\sigma^2_{m_{\ast}} \times |\epsilon|.
\end{equation}
In all the bins, we have $\Delta B$ varying between $\sim 0.0001$ and $\sim 0.0028$ -- therefore the size--mass results are robust against Eddington bias from stellar mass uncertainties.

Malmquist bias near the completeness limit can steepen or flatten the fitted slopes by preferentially including brighter, larger galaxies at fixed mass. We assess this by removing galaxies within $0.5$~dex of the mass completeness limit and refitting. The resulting slope variation is on average $\sim 0.02$, with a maximum of $\sim 0.06$ (for the highest-$z$ linear quiescent fit) -- small enough to leave our conclusions unchanged.

\section{Discussion}
\label{sec:discussion}

Having obtained size--mass and size--redshift relations for our galaxy sample in Section~\ref{sec:results}, we now compare these with results from the broader literature (Section~\ref{subsec:comparisons}), highlight the differences between the different classifications (Section~\ref{subsec:interchange}) and discuss their physical implications (Section~\ref{subsec:interpretations}).

\subsection{Comparisons with other works}
\label{subsec:comparisons}

In contrast to the methods used in this work (see Section \ref{sec:data}), many comparison studies use $UVJ$ classifications, FAST-based stellar masses, and GALFIT or GALFITM measurements corrected to rest-frame 0.5~\textmu m \citep[e.g.][]{vanderwel14, mowla19, nedkova21, ward24}. These differences affect both the galaxy populations being fit and the physical component traced by the measured size, so agreement or disagreement in the fitted parameters in the following subsections should be interpreted in the context of different methodological choices. Table \ref{tab:lit_rev} summarises the methodology, wavelength definition, and fitted parameters for each comparison study; we refer to it throughout this section.

\begin{table*} 
    \centering
    \caption{Posterior median and 68 per cent credible interval for the parameters of the size--mass relation for the quiescent, bulge-dominated, and early-type galaxies using Equation \ref{eqn:dbpl-sm}. Where the parameter is not constrained, we present its < 68th (95th) percentile or its > 32nd (5th) percentile. Dashes are used for parameters where we do not have galaxies to constrain them. To compare with the scale radius $R_0$ in the single-power law fits, instead of reporting $\gamma$, we report $\log_{10}{(R_{\ast}}/{\mathrm{kpc}})$, which is the value of the scale radius $R_{50}$ at $M_{\ast} = 5 \times 10^{10}~\mathrm{M}_{\odot}$ in Equation \ref{eqn:dbpl-sm}.}
    \label{tab:full-fit-dbpl}
    \begin{tabular}{lcccccccc}
    \toprule
     \multirow{2}{*}{sample}& \multicolumn{1}{c|}{redshift} & \multirow{2}{*}{$\log_{10}(R_{\ast}/\mathrm{kpc})$} & \multirow{2}{*}{$\alpha$} & \multirow{2}{*}{$\beta$} &  $\delta$ & \multicolumn{1}{c|}{$\sigma_{\mathrm{int}}$}& \multicolumn{1}{c|}{$N_{\mathrm{gal}}$} & \multicolumn{1}{c|}{$N_{\mathrm{gal}}$} \\
     & bin & & & & dex & dex & $(m_{\ast}\leq 10.3)$ & $(m_{\ast}> 10.3)$ \\ \midrule 
     quiescent: & $[0.0, 0.5)$ & \multicolumn{1}{c|}{$0.39 \pm 0.02$} & \multicolumn{1}{l|}{$0.15 \pm 0.02$} & \multicolumn{1}{c|}{$>1.52~(4.13)$} &  \multicolumn{1}{l|}{$>11.56~(12.08)$} & \multicolumn{1}{c}{$0.15 \pm 0.01$} & 633 & 304 \\ 
    {$\log_{10}(\mathrm{sSFR/\mathrm{yr}^{-1}})$} & $[0.5, 1.0)$ & \multicolumn{1}{c|}{$0.19 \pm 0.01$} & \multicolumn{1}{c|}{$-0.15 \pm 0.05$} & \multicolumn{1}{c|}{$1.07 \pm 0.21$} &  \multicolumn{1}{c|}{$10.76 \pm 0.18$} & \multicolumn{1}{c}{$0.14 \pm 0.01$} & 665 & 1315 \\ 
    {$\leq -11$}
    & $[1.0, 1.5)$ & \multicolumn{1}{c|}{$0.07 \pm 0.02$} & \multicolumn{1}{c|}{$-0.45 \pm 0.19$} & \multicolumn{1}{c|}{$0.98 \pm 0.21$} &  \multicolumn{1}{c|}{$10.45 \pm 0.25$} & \multicolumn{1}{c}{$0.15 \pm 0.01$} & 221 & 898 \\   
    & $[1.5, 2.0)$ & \multicolumn{1}{c|}{$-0.01 \pm 0.02$} & \multicolumn{1}{c|}{--} & \multicolumn{1}{c|}{$0.94 \pm 0.34$} & \multicolumn{1}{c|}{--}&  \multicolumn{1}{c}{$0.15 \pm 0.01$}& 52 & 529 \\ 
    & $[2.0, 3.0]$ & \multicolumn{1}{c|}{$-0.16 \pm 0.04$} & \multicolumn{1}{c|}{--} & \multicolumn{1}{c|}{$>0.71~(0.99)$} & \multicolumn{1}{c|}{--}&  \multicolumn{1}{c}{$0.14 \pm 0.02$} & 11 & 166 \\ \midrule
    bulge-dominated:  & $[0.0, 0.5)$ & \multicolumn{1}{c|}{$0.41 \pm 0.02$} & \multicolumn{1}{l|}{$0.23 \pm 0.01{^\dagger}$} & \multicolumn{1}{c|}{$<4.46~(8.68)$} &  \multicolumn{1}{l|}{$>11.69~(12.43)^{\dagger}$} & \multicolumn{1}{c}{$0.21 \pm 0.01$} & 454 & 205 \\ 
    {$\mathrm{B/T} \geq 0.6$} & $[0.5, 1.0)$ & \multicolumn{1}{c|}{$0.24 \pm 0.02$} & \multicolumn{1}{c|}{$-0.01 \pm 0.03$} & \multicolumn{1}{c|}{$1.70 \pm 1.16$} &  \multicolumn{1}{c|}{$11.35 \pm 0.21$} & \multicolumn{1}{c}{$0.18 \pm 0.01$} & 1298 & 1099   \\ 
    & $[1.0, 1.5)$ & \multicolumn{1}{c|}{$0.19 \pm 0.01$} & \multicolumn{1}{c|}{$-0.10 \pm 0.05$} & \multicolumn{1}{c|}{$0.97 \pm 0.30$} &  \multicolumn{1}{c|}{$10.83 \pm 0.26$} & \multicolumn{1}{c}{$0.18 \pm 0.01$} & 1379 & 1519 \\ 
    & $[1.5, 2.0)$ &\multicolumn{1}{c|}{$0.11 \pm 0.02$} & \multicolumn{1}{c|}{$-0.11 \pm 0.05$} & \multicolumn{1}{c|}{$>1.10~(1.66)$} & \multicolumn{1}{c|}{$>11.04~(11.31)$}&  \multicolumn{1}{c}{$0.19 \pm 0.01$} & 1014 & 1016  \\
    & $[2.0, 3.0]$ & \multicolumn{1}{c|}{$0.03 \pm 0.03$} & \multicolumn{1}{c|}{$-0.11 \pm 0.04$} & \multicolumn{1}{c|}{$>1.33~(4.10)$} & \multicolumn{1}{c|}{$>11.42~(11.96)$}&  \multicolumn{1}{c}{$0.21 \pm 0.01$} & 1348 & 545 \\ \midrule
    early-type: & $[0.0, 0.5)$ &\multicolumn{1}{c|}{$0.42 \pm 0.03$} & \multicolumn{1}{l|}{$0.28 \pm 0.03$} & \multicolumn{1}{c|}{$>0.74~(2.89)$} &  \multicolumn{1}{l|}{$>11.66~(12.37)$} & \multicolumn{1}{c}{$0.21 \pm 0.01$} & 396 & 350 \\ 
    {$n > 2.5$} & $[0.5, 1.0)$ & \multicolumn{1}{c|}{$0.27 \pm 0.01$} & \multicolumn{1}{c|}{$0.08 \pm 0.03$} & \multicolumn{1}{c|}{$>1.15~(1.82)$} &  \multicolumn{1}{c|}{$>11.24~(11.54)$} & \multicolumn{1}{c}{$0.18 \pm 0.01$} & 1189 &1660  \\ 
    & $[1.0, 1.5)$ & \multicolumn{1}{c|}{$0.18 \pm 0.02$} & \multicolumn{1}{c|}{$-0.01 \pm 0.03$} & \multicolumn{1}{c|}{$>1.52~(2.99)$} & \multicolumn{1}{c|}{$>11.40~(11.77)$} & \multicolumn{1}{c}{$0.20 \pm 0.01$} & 1450 & 1290 \\ 
    & $[1.5, 2.0)$& \multicolumn{1}{c|}{$0.13 \pm 0.02$} & \multicolumn{1}{c|}{$-0.06 \pm 0.02$} & \multicolumn{1}{c|}{$>1.38~(3.58)$} & \multicolumn{1}{c|}{$>11.38~(11.88)$}&  \multicolumn{1}{c}{$0.21 \pm 0.01$} & 1247 & 981 \\
    & $[2.0, 3.0]$ & \multicolumn{1}{c|}{$0.08 \pm 0.03$} & \multicolumn{1}{c|}{$-0.04 \pm 0.05$} & \multicolumn{1}{c|}{$>0.68~(3.07)$} & \multicolumn{1}{c|}{$>11.34~(12.08)$}&  \multicolumn{1}{c}{$0.22 \pm 0.01$} & 1948 & 636 \\\bottomrule
    \end{tabular}
\begin{tablenotes}
\setlength{\multicolsep}{0.1cm}
%\begin{multicols}{1}
\item[$\dagger$] $^\dagger$ Bimodal posterior, only the main mode ($\approx 81$~per cent of posterior samples) reported. The secondary mode ($\approx 19$ per cent) pins $\delta$ at the lower prior edge and likewise locates no break within the data.
%\end{multicols}
\end{tablenotes}
\end{table*}

\subsubsection{Size--mass relation}

For the size--mass relation, our results broadly reproduce the established separation between star-forming and quiescent systems. Star-forming galaxies are well described by a single, shallow power law, whereas quiescent systems have a much steeper evolution at the high-mass end than at the low-mass end, being better represented by a double-power law. The star-forming slopes in Table \ref{tab:lit_rev} generally fall within the range spanned by previous work, although the highest-redshift bins are slightly shallower than studies using the rest-frame optical measurements.

The structural splits provide a complementary comparison. Disc-dominated and late-type galaxies occupy the shallow, extended sequence usually associated with star-forming systems, while bulge-dominated and early-type galaxies show steeper high-mass trends. However, the structural and sSFR selections are not interchangeable. The clearest difference is in the intrinsic scatter: when the sample is split by sSFR, $\sigma_{\rm int}$ is broadly similar for star-forming and quiescent galaxies, whereas structural splits produce a stronger separation, with larger scatter for bulge-dominated and early-type systems than for disc-dominated or late-type galaxies.

\begin{table*}
\centering
    \caption{Posterior median and 68 per cent credible interval for the parameters describing the size--redshift relation.}
    \begin{tabular}{lcc}
    \toprule
    sample & $\log_{10}{(B_z~/~\mathrm{kpc})}$ & $\beta_z$ \\
    \midrule
    full & $0.64 \pm 0.02$ & $0.71 \pm 0.07$ \\
    star-forming:  $\log_{10}(\mathrm{sSFR}/\mathrm{yr}^{-1}) > -11$& $0.74 \pm 0.03$ & $0.82 \pm 0.04$ \\
    quiescent, linear: $\log_{10}(\mathrm{sSFR}/\mathrm{yr}^{-1}) \leq -11$ & $0.52 \pm 0.06$ & $1.12 \pm 0.09$ \\
    quiescent, double-power law: $\log_{10}(\mathrm{sSFR}/\mathrm{yr}^{-1}) \leq -11$ & $0.48 \pm 0.07$ & $1.16 \pm 0.22$ \\
    disc-dominated: $\mathrm{B/T} \leq 0.2$ & $0.77 \pm 0.02$ & $0.66 \pm 0.04$ \\
    intermediate: $0.2 < \mathrm{B/T} < 0.6$ & $0.59 \pm 0.03$ & $0.81 \pm 0.07$ \\
    bulge-dominated, linear: $\mathrm{B/T} \geq 0.6$ & $0.49 \pm 0.06$ & $0.74 \pm 0.09$ \\
    bulge-dominated, double-power law: $\mathrm{B/T} \geq 0.6$ & $0.49 \pm 0.06$ & $0.74 \pm 0.17$ \\
    late-type: $n \leq 2.5$ & $0.71 \pm 0.01$ & $0.68 \pm 0.02$ \\
    early-type, linear: $n > 2.5$ & $0.49 \pm 0.03$ & $0.74 \pm 0.09$ \\
    early-type, double-power law: $n > 2.5$ & $0.45 \pm 0.07$ & $0.74 \pm 0.20$ \\
    \bottomrule
    \end{tabular}
    \label{tab:size_redshift_fits}
\end{table*}

Beyond direct evolutionary studies, size--mass parametrisations of the form fitted here can also be used to assign sizes in survey forward models \citep[e.g. GalSBI-SPS;][]{tortorelli25}, where the choice of classification scheme and functional form propagates directly into simulated image-level observables.

\subsubsection{Size--redshift relation}

The size--redshift comparison in Table \ref{tab:lit_rev} shows a wider spread across the literature than the size--mass comparison, but it is largely explained by three methodological choices. The first important difference is the redshift baseline over which a single-power law is fit. Studies extending beyond $z \gtrsim 3$ often recover shallower global $\beta_z$ values than studies restricted to $z < 3.0$, because size evolution appears to flatten at high redshift \citep{ward24, carreira26}. Fitting one power law over a longer baseline can therefore pull the inferred slope downward. This effect is particularly relevant when comparing our $0.0 < z \leq 3.0$ measurements with JWST studies that extend to much higher redshift.

A second difference is population mixing. Full-sample slopes combine populations that have different normalisations and potentially different evolutionary rates, so the measured $\beta_z$ depends both on the intrinsic evolution of each population and on the changing population mix with redshift. In our sample, the full-sample slope is slightly shallower than the star-forming slope, as expected given that the sample is dominated by star-forming galaxies while also including structurally diverse systems. 

The third difference is the wavelength definition of size. Our $R_{50}$ measurements are based on the COSMOS2025 joint NIRCam fits and therefore correspond to a wavelength-averaged observed-frame size, whereas many comparison studies measure or correct sizes to rest-frame $0.5$~\textmu m. As \cite{jia24} show, the inferred $\beta_z$ depends on the wavelength at which size is measured; our approach traces the bulk stellar mass distribution (as described in Section \ref{sec:cosmos2025}), rather than a single rest-frame component, and partly explains the offset from rest-frame optical studies. Using the colour-gradient scaling $\Delta \log{R_{\mathrm{eff}}} / \Delta \log{\lambda} \approx - 0.25$ (\citealp{vanderwel12}; adopted by \citealp{genin25}), the drift of the rest-frame wavelength probed by our joint fit (effective $\lambda_{\mathrm{obs}} \approx 2.5$~\textmu{}m) lowers the fitted $\beta_z$ by $\sim 0.25$ relative to sizes at fixed rest-frame $0.5$~\textmu{}m.

Taken together, these three factors could explain the literature spread without requiring a physical disagreement between studies.

\begin{table*}
\centering
\caption{Comparison of size--mass ($B$ and $\sigma_{\mathrm{int}}$) and size--redshift ($\beta_z$) relations, methodology, and results. For studies with multiple redshift bins, we report the range spanned by the quoted best-fitting values across the relevant redshift interval. Full redshift-dependent fits for this work are given in Tables \ref{tab:full-fit}--\ref{tab:size_redshift_fits}.}
\label{tab:lit_rev}
\begin{threeparttable}
%\scriptsize
\begin{tabular}{lcccccccc}
\toprule
study & $z$ range & min.\ $\log_{10}(M_{\ast}/\mathrm{M}_\odot)$ & rest-frame $\lambda$ & population & $B$ & $\sigma_{\rm int}$~/dex & $\beta_z$\\ 
\midrule
Tudorache et al.\ (this work)\tnote{a,b,c} & $0$--$3$ & & $\sim1$--$5$\textmu m avg. & full & $0.10$--$0.23$ & $0.16$--$0.17$ & $0.71$ \\
& & $7.0$~(SF) & & SF\tnote{d} & $0.11$--$0.26$ & $0.15$--$0.17$ & $0.82$ \\
& & $10.3$~(Q) & & Q\tnote{d} & $0.52$--$0.62$ & $0.12$--$0.15$ & $1.12$ \\
& & & & disc\tnote{e} & $0.18$--$0.27$ & $0.13$--$0.15$ & $0.66$\\
& & & & bulge\tnote{e} & $0.37$--$0.58$ & $0.16$--$0.20$ & $0.74$\\
& & & & late\tnote{f} & $0.15$--$0.25$ & $0.14$--$0.16$ & $0.68$\\
& & & & early\tnote{f} & $0.31$--$0.55$ & $0.16$--$0.22$ & $0.74$\\
\midrule
\citet{carreira26}\tnote{g} & $1$--$15$ & & $0.5$~\textmu m & full & -- & -- & $0.64$ \\
& & & & disc\tnote{e,h} & -- & -- & $1.09$ \\
& & & & bulge\tnote{e,h} & -- & -- & $0.23$ \\
\midrule
\citet{dimauro19}\tnote{i,j,k} & $0$--$2$ & $10.0$ & $0.5$~\textmu m & SF\tnote{l} & $0.08$--$0.33$ & $0.01$--$0.03$\tnote{m} & -- \\
& & & & Q\tnote{l} & $0.14$--$0.48$ & $0.01$--$0.05$\tnote{m} & -- \\
& & & & disc\tnote{h,n} & $0.08$--$0.33$ & $0.16$--$0.18$ & -- \\
& & & & bulge\tnote{h,n} & $0.14$--$0.48$ & $0.16$--$0.18$ & -- \\
\midrule
\citet{genin25}\tnote{b,o,p} & $0$--$6$ & $10.0$ & $\sim0.8$--$5$\textmu m avg. & full & -- & -- & $0.59$ \\
& & & & disc\tnote{q} & -- & -- & $0.61$ \\
& & & & bulge\tnote{q} & -- & -- & $0.87$ \\
\midrule
\citet{gozaliasl25}\tnote{r,s} & $0.08$--$3.7$ & $9.0$ & $\sim0.7$--$1.1$\textmu m & SF (BGG)\tnote{t} & $-0.15$--$0.32$ & $0.19$--$0.34$ & $1.11$ \\
& & & & Q (BGG)\tnote{t} & $0.23$--$0.85$ & $0.15$--$0.42$ & $1.40$\\
\midrule
\citet{ji26}\tnote{u,v} & $0.5$--$5$ & $10.0$ & optical & $M_{\ast}<10^{10.6}\mathrm{M}_\odot$, Q\tnote{l} & -- & -- & $1.1$ \\
& & & & $M_{\ast}>10^{10.6}\mathrm{M}_\odot$, Q\tnote{l} & -- & -- & $1.7$ \\
\midrule
\citet{jia24}\tnote{v,w} & $0$--$3.5$ & $8.0$ & $0.45$~\textmu m & SF\tnote{d} & $0.23$ & -- & $1.04$ \\
& & & $1.0$~\textmu m & SF\tnote{d} & $0.20$ & -- & $1.08$ \\
\midrule
\citet{mowla19}\tnote{i,v,x,y} & $0.1$--$3$ & & $0.5$~\textmu m & full & -- & -- & $0.95$ \\ 
& & $9.0$~(SF) & & SF\tnote{l} & $0.18$--$0.29$ & -- & $1.40$ \\
& & $11.3$~(Q) & & Q\tnote{l} & $0.48$--$0.73$ & -- & $1.09$ \\
\midrule
\citet{nedkova21}\tnote{i,j,k} & $0$--$3$ & $7.5$~(SF) & $0.5$~\textmu m & SF\tnote{l} & $0.17$--$0.24$ & $0.25$\tnote{z} & -- \\
& & $10.3$~(Q) & & Q\tnote{l} & $0.61$--$0.68$ & $0.2$--$0.3$\tnote{z} & -- \\
\midrule
\citet{ormerod24}\tnote{p,v} & $0.5$--$8$ & $9.5$ & optical & full & -- & -- & $0.71$ \\
& & & & $n<2$ & -- & -- & $0.79$ \\
& & & & $n>2$ & -- & -- & $0.44$ \\
\midrule
\citet{varadaraj24}\tnote{aa,ab} & $2.75$--$5.5$ & $9.0$ & F356W & full & -- & -- & $0.60$ \\
\midrule
\citet{ward24}\tnote{i,j,w,ac} & $0.5$--$5.5$ & $9.5$ & $0.5$~\textmu m & SF\tnote{l} & $0.15$--$0.19$ & $0.20$--$0.23$ & $0.63$\\
\midrule
\citet{vanderwel14}\tnote{i,v,y} & $0$--$3$ & $9.0$ & $0.5$~\textmu m & full & -- & -- & $\sim1.1$ \\
& & & & SF\tnote{l} & $0.18$--$0.25$ & $0.16$--$0.19$ & $\sim0.75$\\
& & & & Q\tnote{l} & $0.71$--$0.79$ & $0.10$--$0.14$ & $\sim1.48$\\
\midrule
\citet{yang25}\tnote{r,s} & $2$--$10$ & $9.0$ & $0.5$~\textmu m & SF\tnote{t} & $-0.13$--$0.37$ & $0.17$--$0.28$ & $1.21$ \\
& & & & Q\tnote{t} & $0.52$--$0.78$ & $0.16$--$0.25$ & $0.81$\\
\midrule
\citet{yu26}\tnote{b,r} & $0.5$--$3$ & $10.0$ & $1.22$~\textmu m & SF\tnote{l} & -- & -- & $0.92$ \\
& & & & Q\tnote{l} & -- & -- & $1.34$\\
\bottomrule
\end{tabular}
 
\begin{tablenotes}\scriptsize
\setlength{\multicolsep}{0.1cm}
\begin{multicols}{2}
\item[a] Stellar masses from \texttt{pop-cosmos} SED fits \citep{thorp24, thorp25b}.
\item[b] Sizes based on \textsc{SourceXtractor++} \citep{bertin20}.
\item[c] Sizes based on a 4-band joint fit.
\item[d] SF/Q split based on sSFR.
\item[e] Disc/bulge split based on $\mathrm{B/T}$.
\item[f] Late-/early-type split based on S\'{e}rsic index, $n$.
\item[g] Sizes based on \textsc{pysersic} \citep{pasha23}.
\item[h] Sizes for disc/bulge galaxies based on their disc/bulge component.
\item[i] Stellar masses from \textsc{FAST} \citep{kriek09}.
\item[j] Sizes based on \textsc{GALFITM} \citep{vika13, haussler13, haussler22}.
\item[k] Sizes based on a 7-band joint fit.
\item[l] SF/Q split based on $UVJ$ diagram.
\item[m] The \cite{dimauro19} $\sigma_{\rm int}$ values quoted for integrated mass--size fits are lower than their component-based bulge/disc scatters as their fitting method is weighted by the inverse number density and allowing 1~per cent outliers, producing artificially tight fits.
\item[n] Disc/bulge component fit; whole-galaxy classification by B/T.
\item[o] Sizes based on a joint fit of 7--9 bands (depending on the field).
\item[p] Stellar masses from \textsc{EAZY} \citep{brammer08}.
\item[q] Disc/bulge split based on combined S\'{e}rsic-index and B/T criteria (disc: $n<4$ and B/T$<0.5$; bulge: $n>1$ and B/T$>0.5$).
\item[r] Stellar masses from \textsc{LePhare} \citep{arnouts99, ilbert06, ilbert09}.
\item[s] Sizes based on band-by-band single-S\'{e}rsic fits with \textsc{Galight} \citep{ding20}, using a redshift-dependent NIRCam band to probe the quoted rest-frame wavelength.
\item[t] SF/Q split based on the rest-frame $NUV-r-J$ diagram combined with a redshift-dependent sSFR criterion.
\item[u] Stellar masses from \textsc{Prospector} \citep{johnson21} SED fits.
\item[v] Sizes based on \textsc{GALFIT} \citep{peng02}.
\item[w] Stellar masses from a combination of SED fits.
\item[x] Stellar masses based on \citet{bruzual03} templates.
\item[y] Sizes based on single-band \textsc{GALFIT} fits.
\item[z] Intrinsic scatter estimated from root mean square (RMS) error.
\item[aa] Stellar masses from \textsc{Bagpipes} \citep{carnall18}.
\item[ab] Sizes based on \textsc{PyAutoGalaxy} \citep{nightingale23}.
\item[ac] Sizes based on a 13-band joint fit.
\end{multicols}
\end{tablenotes}
\end{threeparttable}
\end{table*}

\begin{figure*} 
    \centering
    \captionsetup{justification=centering}
    \includegraphics[width=1\linewidth]{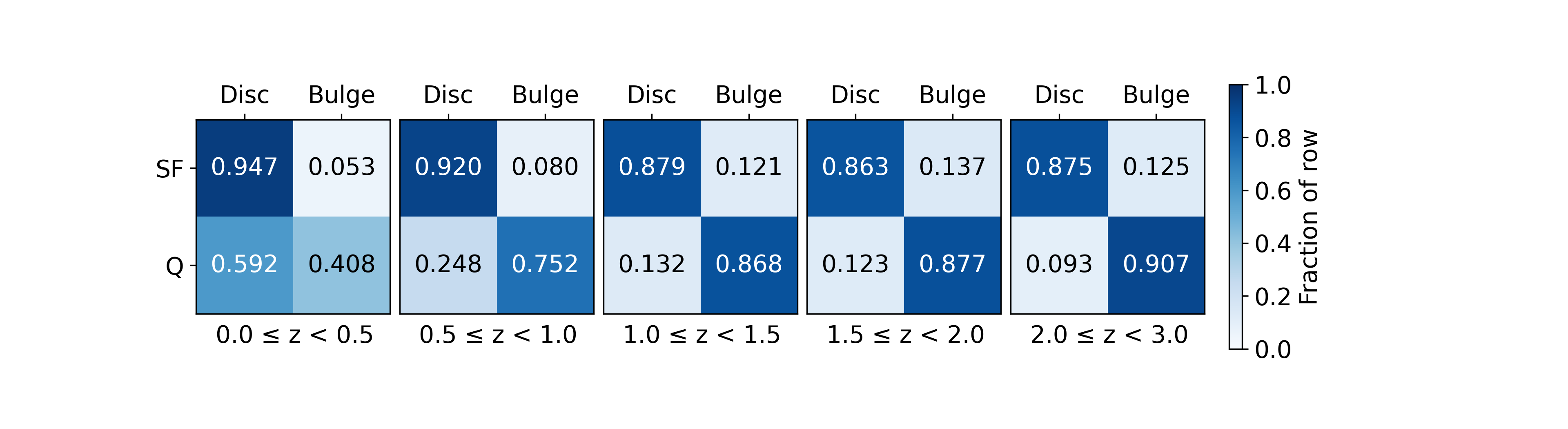} 
    \caption{Contaminant fraction between the sSFR and morphology classification -- i.e.\ how many disc- ($\mathrm{B/T}\leq 0.2$) and bulge-dominated ($\mathrm{B/T}\geq0.6$) galaxies are in the star-forming and quiescent samples compared to the full samples -- in each of the five redshift bins.}
    \label{fig:sf-q-b-d-contamination}
\end{figure*}

\subsection{Morphology markers and sSFR are not interchangeable}
\label{subsec:interchange}

The results in Section \ref{sec:results} show that sSFR-based and morphology-based classifications are not interchangeable: disc-dominated galaxies show steeper, tighter size--mass relations than the full star-forming sample (slope $B$ offset of $0.01$--$0.08$ across redshift bins, $\sigma_{\mathrm{int}} \sim 0.13$--$0.15$ vs $0.15$--$0.17$~dex); late-type galaxies show the same pattern relative to the star-forming sample, though with a smaller offset. Quiescent and bulge-dominated galaxies likewise differ: their pivot masses in the double-power law fits are offset by $\sim 0.5$~dex (quenching precedes morphological transformation, see Section \ref{subsubsec:q-vs-b}), and their linear slopes differ from those of early-type galaxies at the same redshift. Whilst we are able to isolate a transition mass in the double-power law for the bulge-dominated population, this is not true for the early-type population. Bulge-to-total ratio should not be used as a proxy for star-formation activity, and S\'{e}rsic index should not be used as a one-to-one proxy for either. 

Figure \ref{fig:sf-q-b-d-contamination} quantifies the contamination between the star-forming/quiescent and bulge-to-total ratio classification schemes. The fraction of bulge-dominated galaxies within the star-forming population increases steadily with redshift, rising from 5~per cent to 14~per cent across successive bins. Including these contaminants flattens the star-forming size--mass slope $B$ and increases its scatter relative to the disc-dominated sample. In contrast, contamination of the quiescent population by disc-dominated galaxies is especially pronounced at low redshift, where 59~per cent of quiescent systems at $z<0.5$ are disc-dominated, before declining to 9~per cent at higher redshift. This indicates that at low redshift the majority of quiescent galaxies have ceased forming stars without first becoming bulge-dominated. Our findings at low redshift extend those at $z \approx 0$ by \citet{stephenson24}, whose SDSS central-galaxy sample shows that massive galaxies just above the quiescent threshold are structurally disc-like -- low S\'{e}rsic index and low central density -- with declining recent star-formation histories (their fig. 19). We note that their subsamples are defined relative to the star-forming main sequence rather than by an absolute sSFR cut, so the comparison holds only at the high-mass end ($m_{\ast} \geq 10.5$) where the two schemes overlap. Taken together, these results show that the population feeding the high mass quiescent class at late times is structurally unprocessed.

\begin{figure*} 
    \centering
    \captionsetup{justification=centering}
    \includegraphics[width=1\linewidth]{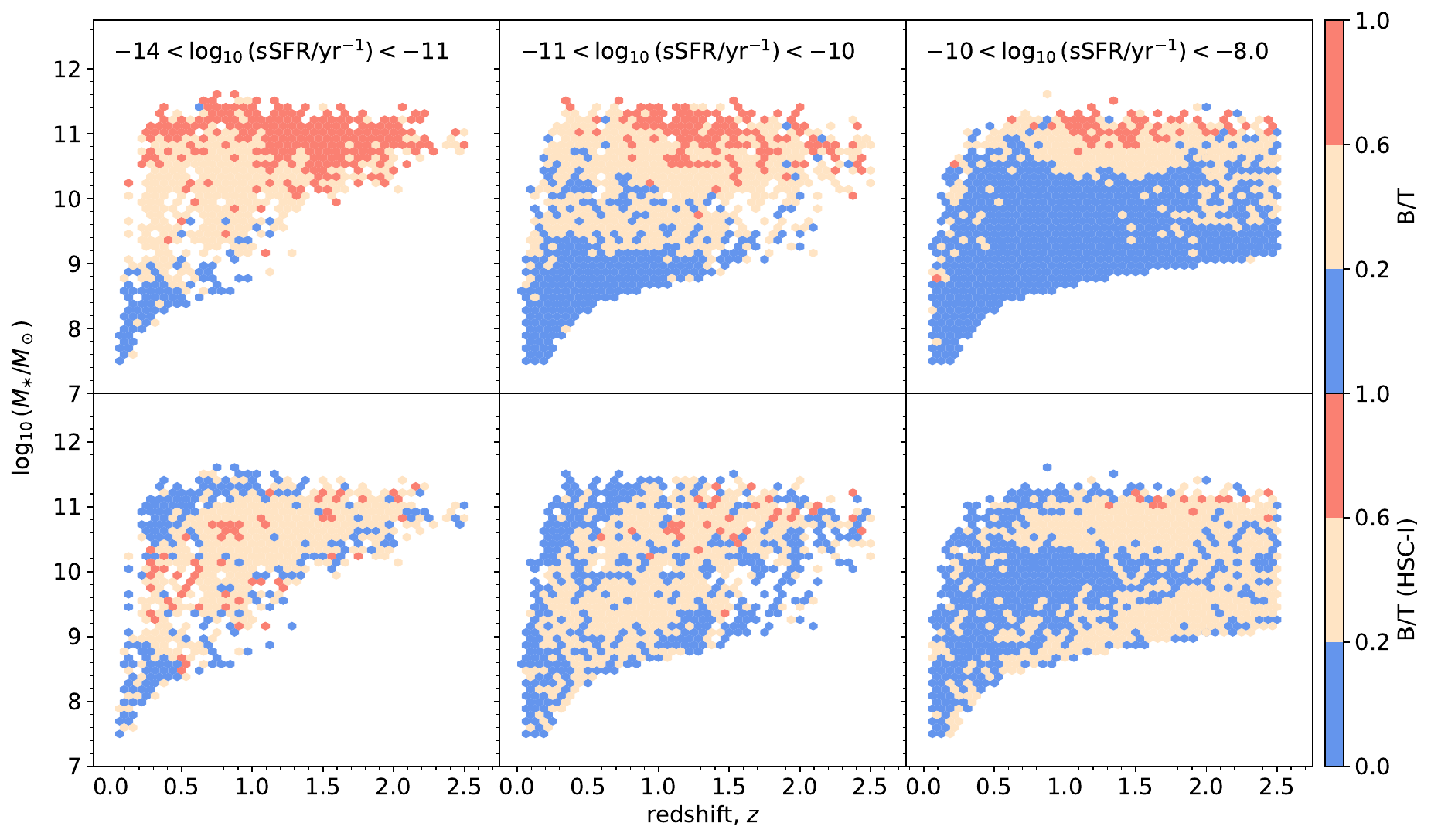} 
    \caption{Stellar mass versus redshift, colour-coded by bulge-to-total ratio measured from F115W for the galaxies between $0.0 < z < 2.0$ and F150W for the galaxies between $2.0 \leq z \leq 3.0$ (\textit{top row}) and HSC-I (\textit{bottom row}), shown separately for three sSFR bins. The space-based B/T recovers a clear correlation between B/T, stellar mass, and redshift that is washed out in the ground-based measurement, demonstrating that the morphology-dependent trends identified in this work require space-based resolution.}
    \label{fig:ms-z-bt-ssfr-bins}
\end{figure*}

\begin{figure}
    \centering
    \includegraphics[width=\columnwidth]{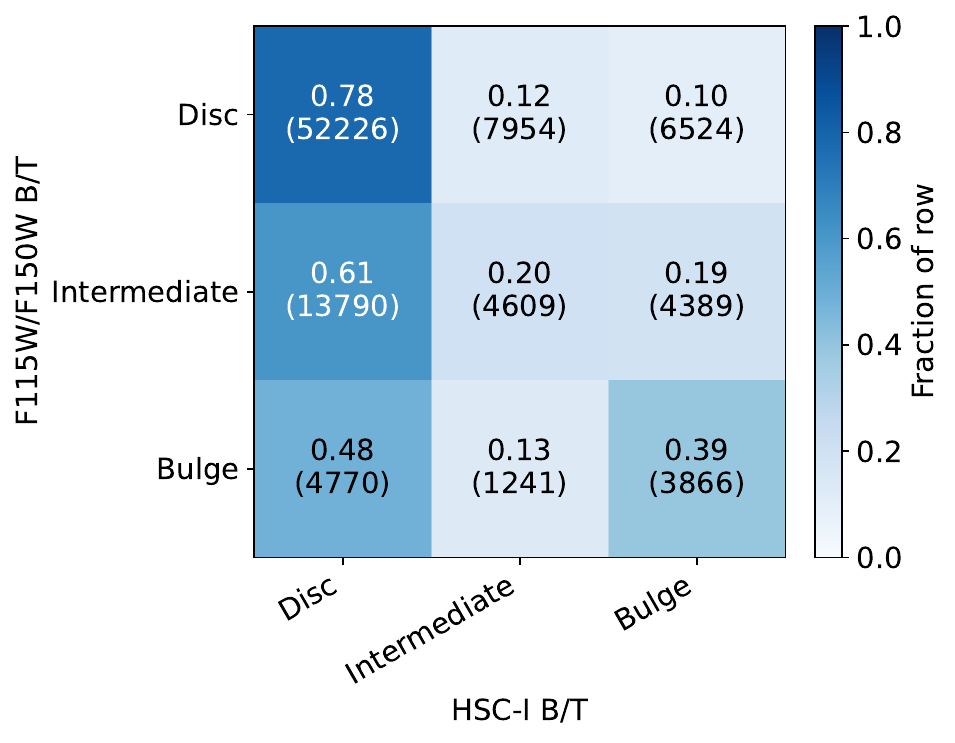}
    \caption{Confusion matrix between the morphology classification from the space-based B/T measurements (rows) and the ground-based B/T measurements (columns), illustrating the level of misclassification of space-based B/T F115W/F150W measurements by ground-based HSC-I determinations. Each row is normalised to unity, with galaxy counts in parentheses.}
    \label{fig:confusion-matrix}
\end{figure}

This structural diversity within each sSFR class is illustrated in Fig. \ref{fig:ms-z-bt-ssfr-bins}, where the quiescent bin contains galaxies spanning all three B/T classes, demonstrating that quiescence does not map cleanly onto a single structural regime. Space-based measurements make this overlap more apparent, as the improved resolution resolves structural classes that ground-based imaging blurs together. To quantify how much of the space-based classification survives at ground-based resolution, Figure \ref{fig:confusion-matrix} compares, galaxy by galaxy, the B/T class (disc-dominated, intermediate or bulge-dominated; B/T $\leq 0.2$, $0.2$--$0.6$ and $\geq 0.6$) assigned from the F115W/F150W and HSC-I measurements. The two classifications agree for $61$ per cent of galaxies, but because HSC-I assigns $71$ per cent of galaxies to the disc class, random assignment with the same class frequencies would already agree for 52 per cent; the agreement beyond chance is small (Cohen's $\kappa = 0.18$; \citealt{cohen60}). Only $39$ per cent of galaxies that are bulge-dominated in the space-based measurement are recovered as such from the ground, while $48$ per cent are assigned to the disc class. The ground-based B/T therefore does not reproduce the space-based classes, which is why the B/T trends with stellar mass, redshift and sSFR are washed out in the bottom panels of Fig.~\ref{fig:ms-z-bt-ssfr-bins}. This has direct implications for any approach that attempts to learn morphology–galaxy property correlations from seeing-limited data, including recent multimodal foundation models \citep{parker24, parker25}.

The same effect occurs for the B/T selection versus the S\'{e}rsic index selection. Whilst both are morphology markers, they capture different stages, as there is significant scatter between the two \citep{gadotti09, yang26}. Specifically, the early-type ($n > 2.5$) sample captures a broader range of B/T values than the bulge-dominated sample, which dilutes the structural homogeneity needed for the double-power law to resolve a transition mass. We discuss this in more depth in Appendix \ref{bt-vs-n}.

Because the single S\'{e}rsic size $\hat{R}_{50}$ and the S\'{e}rsic index $\hat{n}$ are derived from the same joint NIRCam fit, their measurement errors are covariant, and splitting by $\hat{n}$ could in principle imprint a spurious size-morphology correlation. Several features of our results argue against this. First, covariant size--$\hat{n}$ errors would tighten the size--mass relation within S\'{e}rsic-index-selected subsamples rather than broaden it -- but we observe the opposite: early-type  galaxies show larger $\sigma_{\rm int}$ than late-type and disc-dominated galaxies (Tables \ref{tab:bt_cut_fit}--\ref{tab:full-fit-dbpl}). Second, our strongest morphological result -- the pivot mass in the double-power law -- is driven by the B/T split, where B/T is measured from a separate per-band bulge+disc decomposition (separate from the single S\'{e}rsic fit, see Section \ref{sec:cosmos2025}).  The  $\sigma_{\rm int}$ ordering also holds in the B/T split, corroborating the S\'{e}rsic-based result independently of the $\hat{n}$--$\hat{R}_{50}$ covariance. Third, the central comparison of this work is between the sSFR split and the morphology splits, and the sSFR classification carries no morphological information from the COSMOS2025 size catalogue. Measurement covariance within that catalogue therefore cannot generate the contrast we find between the sSFR- and morphology-based relations.

\begin{figure*} 
    \centering
    \captionsetup{justification=centering}
    \includegraphics[width=1\linewidth]{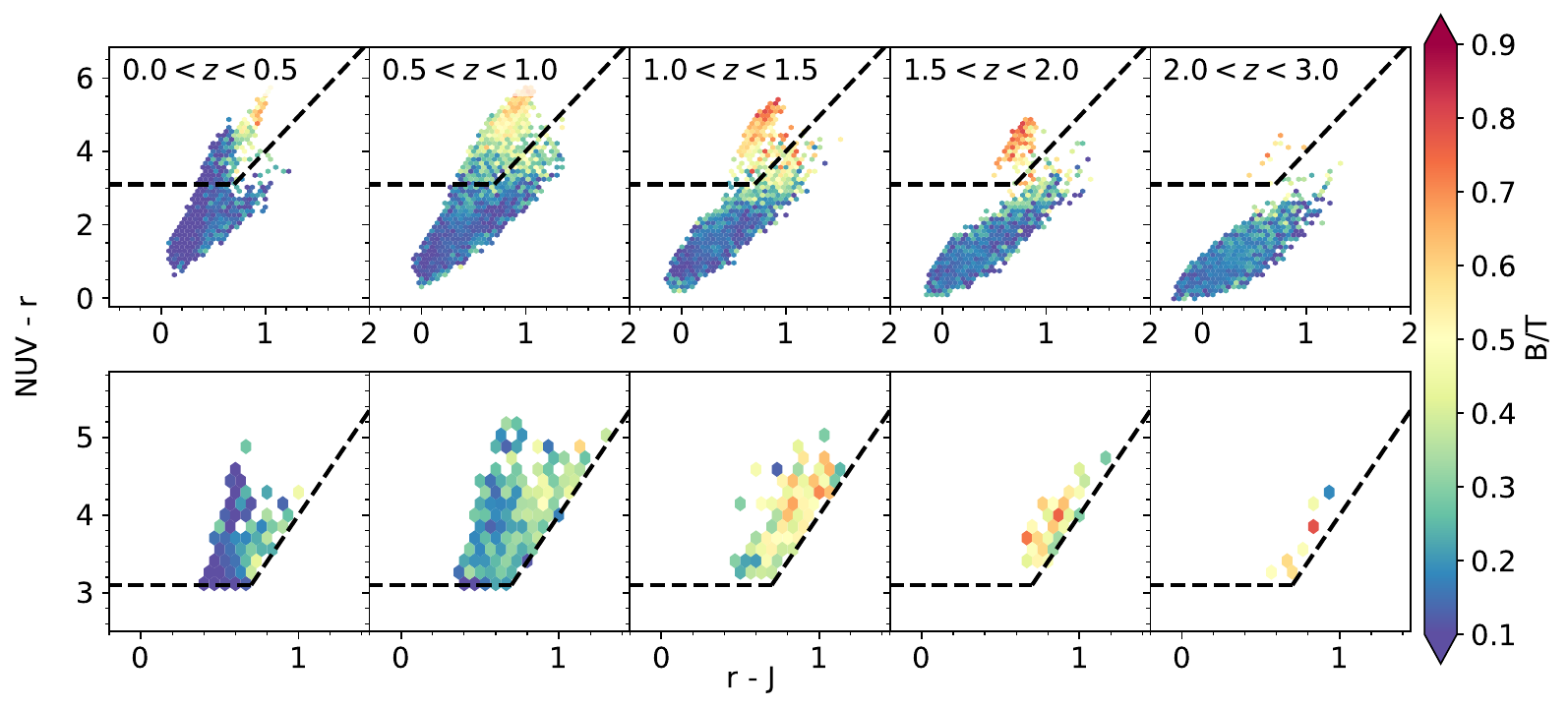} 
    \caption{Rest-frame $NUVrJ$ colour-colour diagram, colour-coded by the bulge-to-total ratio for the whole sample (\textit{top}) and for the galaxies classified as quiescent by the $NUVrJ$ criterion, but classified as star-forming by their sSFR ($\mathrm{sSFR} > 10^{-11} \mathrm{yr}^{-1}$, \textit{bottom}). The black line shows the separation between star-forming and quiescent galaxies according to the rest-frame $NUVrJ$ colour criterion.}
    \label{fig:ms-z-bt-nuvrJ-f115w}
\end{figure*}

Finally, as discussed in Section \ref{subsec:classification}, we consider the use of the $NUVrJ$ selection criterion or the $UVJ$ criterion instead of sSFR to define quiescence (see \citealt{leja19_uvj} for a more in-depth discussion about the pitfalls of $UVJ$ colour selection). The resulting quiescent sample is contaminated by star-forming galaxies. To explore how this affects the correlation with the morphological markers, we investigate the $NUVrJ$ diagram of our selected sample. We define the $NUVrJ$ criterion for quiescent galaxies following \cite{ilbert13} as:
\begin{equation}
(NUV - r) > 3 \, (r - J) + 1~\mathrm{and}~(NUV - r) > 3.1.
\end{equation}
We plot the $NUVrJ$ plane colour-coded by B/T (top panel of Figure \ref{fig:ms-z-bt-nuvrJ-f115w}). In the bottom panels of Figure \ref{fig:ms-z-bt-nuvrJ-f115w}, we show galaxies which are classified as quiescent by the $NUVrJ$ criterion, but are considered star-forming by the sSFR inferred from \texttt{pop-cosmos}. As can be seen, at lower redshifts, these galaxies are disc-dominated, showing that the $NUVrJ$ selection is susceptible to contamination from low-mass disc-dominated star-forming galaxies.

Taken together, Figures \ref{fig:sf-q-b-d-contamination}--\ref{fig:ms-z-bt-nuvrJ-f115w} show that the three classification schemes -- sSFR, B/T, and S\'{e}rsic index -- are not equivalent because they capture different evolutionary stages of the same processes.

\subsection{Physical interpretation}
\label{subsec:interpretations}
The quantitative differences between the classification schemes point to a two-stage picture of galaxy evolution in which star-formation cessation and structural transformation are partially decoupled, driven by different physical processes on different timescales. In this section, we discuss these processes and their implications on galaxy evolution.

\subsubsection{Intrinsic scatter as a probe of the galaxy-halo connection}
\label{subsubsec:intrinsic-scatter}

We find that $\sigma_{\mathrm{int}}$ is independent of sSFR, but has a dependence on either B/T or S\'{e}rsic index, which further emphasises the point that star-formation activity and morphology markers should be considered separately. This has important implications for the galaxy-halo connection, and it follows directly from simple analytic disc-formation models. If galaxy sizes are set primarily by the angular momentum of their host dark matter haloes, then the scatter in the size–mass relation should reflect the dispersion in halo spin \citep{mo98}. Numerical simulations find a $\log_{10}$ dispersion in halo spin of $\sim 0.2$--$0.25$~dex \citep{maccio08, kravtsov13, burkert16}, which is similar to the value of the intrinsic scatter of the size--mass relation ($\sigma_{\mathrm{int}} \sim 0.12$--$0.25$~dex, depending on the split). The fact that $\sigma_\mathrm{int}$ tracks morphology but not sSFR therefore implies that it is the structural configuration of a galaxy -- not its star-formation activity -- that connects to halo occupation. Disc-dominated galaxies, whose sizes are set by angular momentum conservation during gas cooling, retain a tighter memory of halo spin; bulge-dominated systems, whose sizes have been modified by secular processes and early quenching, show larger scatter because these processes partially erase the halo spin signal. This difference is supported by \cite{lagos17}, who find in EAGLE that the scatter in the stellar specific angular momentum at fixed mass is strongly correlated with morphological proxies across the full galaxy population. They identify two channels producing low-angular-momentum, high-scatter early-type systems at $z = 0$: mergers and early star-formation quenching, the latter even in the absence of significant merging.

The fact that the intrinsic scatter $\sigma_{\mathrm{int}}$ does not depend on sSFR is consistent with other observational studies \citep{vanderwel14, dimauro19}. We find a difference in $\sigma_\mathrm{int}$ between disc-dominated and intermediate/bulge-dominated systems in the B/T split with a clear ordering, stronger than the near-flat B/T trend in \cite{abdullah25}. We also find a small difference between the late and early-type systems, similar to \cite{shen03}. However, \cite{shen03} do not treat the intrinsic scatter $\sigma_{\rm int}$ as a free parameter, but estimate the log-normal dispersion $\sigma_{\ln{R}}$ independently in each luminosity bin and then fit its mass dependence with a smooth parametric function, finding that scatter is itself a function of mass.

\subsubsection{Galaxy growth across time for the star-forming population}
\label{subsubsec:sf-growth}

The single-power law form of the star-forming size--mass relation implies that a single growth mechanism operates across the full mass range: continuous in-situ star formation, which builds stellar mass and extends galaxy sizes gradually at all masses \citep{vanderwel14, mowla19, nedkova21}. This form carries a physical prediction. If in-situ growth dominates, then star-forming galaxies should grow their sizes smoothly with cosmic time, and galaxies that will eventually quench should already be structurally offset from their star-forming contemporaries before quenching occurs, not because of any post-quenching process but because compact progenitors were always less extended.

The in-situ growth regime for the star-forming population is also supported by simulations. \cite{genel18} track main progenitors in TNG100 and show that main-sequence galaxies grow their sizes continuously with cosmic time, whereas galaxies that will quench show much weaker size growth before quenching. They further show that galaxies which eventually quench were already smaller than their star-forming contemporaries before quenching occurred. This provides a natural explanation for the population-wide size offset between our quiescent and star-forming samples visible in Tables \ref{tab:full-fit} and \ref{tab:full-fit-sf} -- $\log_{10}{(R_0/\mathrm{kpc})}$ is $0.25$--$0.42$~dex lower for quiescent galaxies in every redshift bin. Our per-redshift fits cannot trace individual progenitors to confirm this directly; but our data are consistent with the mechanism proposed in \cite{genel18}. \cite{pillepich19} demonstrate that the star-forming component remains structurally extended and dynamically disc-like across a wide mass and redshift range, consistent with star formation contributing efficiently to outward size growth. \cite{furlong17} further show that size growth can be increased by internal stellar redistribution, with compact progenitors experiencing additional growth through renewed star formation.

The size–redshift evolution of star-forming galaxies reinforces this picture. Star-forming discs grow gradually through continued gas accretion regulated by dark matter halo angular momentum, producing moderate size evolution ($\beta_z \sim 0.82$) that broadly tracks halo virial radius growth \citep{huang17, nedkova21, miller26}. The physical contrast with quiescent galaxies -- whose steeper size–redshift slope ($\beta_z \sim 1.12$) reflects accelerated size growth through mechanisms that operate after star formation has ceased -- points directly to the two-phase picture the functional forms encode: gradual, halo-regulated growth for star-forming galaxies, and a distinct regime for quenched systems in which other processes increasingly dominate. What drives that accelerated growth, and what sets the mass scale at which the transition occurs, are the subject of the following section.

\subsubsection{Two-phase evolution and decoupling of quiescence and bulge-formation}
\label{subsubsec:q-vs-b}

The double-power law for quiescent (and bulge-dominated) galaxies reflects a mass-dependent transition between two regimes, split by a pivot mass: lower-mass quiescent galaxies are predominantly recently quenched systems whose sizes remain close to their pre-quenching values, whilst higher-mass quiescent galaxies, which lie on the steep slope, reflect the accumulated size growth of galaxies that have been quenched for longer. This mass-dependent picture is also supported observationally by \cite{ji26}, who study massive quiescent galaxies in JADES and find that their size evolution depends on stellar mass. In their sample, lower-mass quiescent galaxies evolve approximately as $(1+z)^{-1}$, while more massive quiescent galaxies evolve more steeply, approximately as $(1+z)^{-1.7}$, above $M_{\ast} \sim 10^{10.6}\,{\rm M_\odot}$. This transition mass is close to the quiescent pivot mass recovered in our double-power law fits. The slight upturn of the relation at the intermediate redshifts is also reported by \cite{miller26}, who attribute it to two distinct classes of quiescent galaxies formed through these two mechanisms. The physical origin of the pivot mass itself -- the quenching threshold -- is discussed in Section \ref{subsubsec:agn}.

The growth mechanism responsible for the steep high-mass slope remains debated. Minor mergers have long been invoked as the dominant channel, depositing accreted stars preferentially at large radii and inflating effective radii without significantly increasing central mass densities \citep{hopkins10, vandokkum10}. However, this picture has been challenged by more recent simulation work. \cite{quai21} show that recently merged galaxies in IllustrisTNG are less likely to undergo rejuvenation than mass-matched controls, suggesting that the merger channel has been overstated in earlier work that lacked the statistical leverage to test its sufficiency directly. \cite{casimiro26}, tracking over $11\,000$ central galaxies in TNG, further find that mergers are neither necessary nor sufficient to explain the structural growth of quiescent systems: only $\sim 3$ per cent of major mergers are followed by quenching within a Gyr, and once stellar mass is controlled for, there is no excess of mergers in the histories of the quenched galaxies.

The environmental dependence of these transitions offers a complementary perspective: if quenching and structural transformation are sequentially ordered by internal processes, environment should modulate the pathway without shifting the mass threshold at which it begins. \cite{gentile25}, using a mass-complete sample of nearly one million galaxies at $0.25 < z < 1$ from the Euclid Q1 data, find that the ordering of quenching and morphological transformation reverses with local density. In the field, secular bulge growth on the main sequence precedes quenching: star-forming bulge-dominated systems represent up to 25~per cent of field galaxies above $M_{\ast} \sim 10^{10.5}\mathrm{M}_{\odot}$ at $z \sim 0.75$--$1.00$, declining to $\sim 15$~per cent by $z \sim 0.25$. In dense environments the sequence inverts -- quiescent disc-dominated galaxies, negligible in the field (<5~per cent at $0.75 < z < 1$), reach fractions up to 20~per cent in over-dense regions with abundance increasing toward lower redshift, consistent with environmental stripping quenching star formation in disc-dominated satellites before any structural transformation occurs. \cite{ghaffari26} extend this picture to higher redshift and denser systems, studying the 25 richest COSMOS-Web groups out to $z = 3.65$. Even close to the group centres, late-type galaxies outnumber early-type galaxies, and across all 25 groups the red sequence contains more late-type galaxies than early-type galaxies, directly showing that star formation is suppressed in disc-dominated systems before the disc is destroyed. There is a mass-dependence: high-mass group galaxies ($m_{\ast} > 10.5$) transition from late-type-dominated quiescence at $z \sim 3.5$ to early-type-dominated at $z \sim 0.5$, while intermediate-mass members ($9.5 < m_{\ast} < 10.5$) show far milder evolution and low-mass members remain predominantly star-forming at all redshifts.

Group-central galaxies show a tighter coupling. \citet{gozaliasl25} find that, among $\sim 1700$  brightest group galaxies (BGGs) in COSMOS-Web at $0.08 < z < 3.7$, the early-type fraction\footnote{We note that the disc/bulge classification here is done via a S\'{e}rsic index cut at $n=2.5$, equivalent to our early/late-type dichotomy, not on bulge-to-total ratio.} increases progressively from star-forming through transitioning to quiescent BGGs, reaching $\sim 80$~per cent of the quiescent population. The contrast with the group members of \citet{ghaffari26}, whose red sequence remains late-type dominated, indicates that the morphological pathway following quenching depends on the position within the halo. Using the same COSMOS-Web catalogue, \citet{toni26} trace the build-up of the quiescent population and red sequence in these groups, finding steady assembly from $z \sim 1.5$--$2$, earlier in richer systems.

The mass scale above which the environmental trends of \citet{gentile25} and \citet{ghaffari26} are reported, $m_{\ast} \sim 10.5$, is consistent with our quiescent pivot $\delta_{\rm Q} = 10.7 \pm 0.2$ although in both cases it is an analysis threshold rather than a measured transition. In this analysis we do not consider the environment, and hence the question of whether  $\delta_{\rm Q}$ itself is environment-dependent is not tested here. The $\sim 0.5$~dex offset between $\delta_{\rm Q}$ and $\delta_{\rm B}$ is measured for the COSMOS-Web population as a whole.

The ordering implied by $\delta_{\rm Q} < \delta_{\rm B}$ -- quenching first, structural transformation later -- is supported by \cite{park22}, who show in TNG50 that quenched discs are subsequently heated into spheroids over the quiescent period -- a structural transformation that operates after star formation has ceased. This provides a route by which a galaxy crosses the sSFR threshold first and becomes structurally early-type later, consistent with the ordering we observe. Their second pathway -- galaxies that merge into spheroids before quenching -- operates preferentially above $M_{\ast} \sim 10^{11}~\mathrm{M}_{\odot}$, where our quiescent sample is sparse and hence does not provide constraining power. Their morphological classification is kinematic (disc-to-total decomposition by angular momentum) rather than photometric like B/T, so the comparison is not one-to-one. However, they show that their D/T trends track S\'{e}rsic index trends with mass (their fig. 1c). Given the known correlation between S\'{e}rsic index and B/T -- broad but systematic at these masses (\citealt{gadotti09}; Appendix \ref{bt-vs-n}) -- the disc-heating result is qualitatively consistent with the structural transformation inferred from the B/T split.

Taken together, the simulation and observational evidence point to a picture in which the quiescent pivot mass marks a genuine physical threshold set by internal processes, while the subsequent structural evolution into bulge-dominated morphologies proceeds through a combination of disc-heating and secular evolution on timescales long enough to explain the $\sim 0.5$~dex offset between $\delta_{\rm Q}$ and $\delta_{\rm B}$. We stress that the pivot-mass offset is robustly constrained only over $0.5 \leq z < 1.5$, where both $\delta_\mathrm{Q}$ and $\delta_\mathrm{B}$ are located (Section \ref{subsec:size-mass}); the highest-redshift quiescent bins contain few low-mass galaxies ($N = 52$ and $11$ at $1.5 \leq z < 2.0$ and $2.0 \leq z \leq 3.0$; Table \ref{tab:full-fit-dbpl}) and leave $\delta$ unconstrained. The two-phase interpretation rests on the well-constrained intermediate-redshift range, not on a uniform trend across all five bins. What sets the quenching threshold itself is the subject of the following section.

\subsubsection{Linking AGN feedback to quiescence}
\label{subsubsec:agn}

Three mass scales bracket the quenching-to-structural-transformation sequence. We will now discuss how each coincides with a scale at which AGN feedback changes character in simulations -- (i) at $\delta_{\rm Q} \sim 10.7$ feedback becomes able to offset cooling and quenching begins; (ii) at $\delta_{\rm B} \sim 11.1$ AGN-driven baryonic redistribution peaks, enabling disc-heating into a bulge; and (iii) above $M_{\rm crit} \sim 10^{13.5-14}~\mathrm{M}_{\odot}$ (halo mass) heating can no longer prevent re-accretion and the break washes out.

The significance of the first scale is reinforced by \cite{deger25} who, within the \texttt{pop-cosmos} framework, map AGN activity across the stellar mass--sSFR plane. They find that the AGN (infrared torus) bolometric luminosity fraction peaks at $f_\mathrm{AGN} \sim 0.1$--$0.5$ specifically in galaxies approaching the quenching transition -- at stellar masses $m_{\ast} \sim 10.5$ and stellar mass doubling times of $3$--$10$ Gyr -- before declining by one to two orders of magnitude once galaxies reach full quiescence (see their section~$7.3$ and figure~$15$). The AGN activity therefore peaks not just at the right mass scale but at the right evolutionary stage: in galaxies that are still forming stars but at suppressed rates, caught in the act of quenching.

The quiescent pivot mass $\delta_{\rm Q}$ corresponds to a halo mass $M_{200} \sim 10^{12}\,\mathrm{M_\odot}$ via the FLAMINGO stellar-to-halo mass relation (\citealp{schaye23}, their fig.~9). This is the scale at which AGN feedback becomes effective: \citet{bower17} show that above this halo mass the central black hole can offset radiative cooling. The bulge-dominated pivot $\delta_{\rm B}$ maps onto $M_{200} \sim 10^{12.7}\,\mathrm{M_\odot}$, a $0.7$~dex offset in halo mass. \citet{lucie25} show that in FLAMINGO, the efficiency of baryonic redistribution by AGN feedback calibrated to observed gas fractions peaks at $M_{200} \sim 10^{12.8}~\mathrm{M}_{\odot}$ ($m_{\ast} \sim 11.1$), coinciding with $\delta_B$. At this halo mass, feedback has maximally redistributed the circumgalactic gas, altering the gravitational potential in which the disc sits and making it susceptible to heating and thickening. The disc-heating pathway identified by \citet{park22} -- in which quenched discs are secularly heated into spheroids -- therefore provides a plausible reshaping mechanism; the coincidence with \citet{lucie25}'s finding on the baryonic redistribution peak explains why it becomes effective at that mass scale rather than any other. Mergers are not required: as discussed in Section~\ref{subsubsec:q-vs-b}, recent simulation work finds them neither necessary nor sufficient for the structural transformation of quiescent galaxies.

Recent work using the FLAMINGO simulations by \cite{peiris26} and \cite{pontzen26} has suggested that above a higher halo mass $M_\mathrm{crit} \sim 10^{13.5\text{--}14}\,\mathrm{M_\odot}$ ($m_{\ast} \sim 11.2\text{--}11.5$), the post-shock entropy of accreting gas exceeds the entropy ceiling that AGN heating can maintain in the circumgalactic medium. This opens the possibility of massive galaxies rejuvenating due to re-accreting gas. In this scenario, quenching is no longer reliably maintained, so the sharp break between the low-mass and high-mass quiescent regimes should wash out. This is consistent with our fits at $z < 0.5$, where the double-power law pivot is unconstrained ($\delta_{\rm Q} > 11.56$~dex, Table~\ref{tab:full-fit-dbpl}) and the quiescent population shows broad structural diversity -- an unconstrained pivot is consistent with a dissolving break in the fit. A smoothly rising, break-free quiescent size–mass relation is also found at $z \approx 0$ by \citet[][their fig. 3]{stephenson24}. Their quiescent relation shows no located break below $\sim 10^{11.3}~\mathrm{M}_{\odot}$ and converges onto the star-forming relation at high mass. This scenario at the most massive end of galaxy evolution is further supported by \citet{gentile26}, who report two massive star-forming disc galaxies at $z \sim 0.75$, each the brightest galaxy of a group with $M_{200} \sim 10^{13.8}\,\mathrm{M_\odot}$. Both retain molecular gas reservoirs of $\sim 10^{10.3}\,\mathrm{M_\odot}$ and main-sequence star-formation efficiencies -- properties consistent with gas re-accretion above $M_\mathrm{crit}$\footnote{\citet{gentile26} interpret the extended gas reservoirs as debris from a wet merger with a gas-rich satellite, but this interpretation sits in tension with the IllustrisTNG results of \citet{quai21} and \citet{casimiro26} discussed in Section~\ref{subsubsec:q-vs-b}.}.

The coincidence of these scales could be explained by a picture in which the AGN feedback operates across three distinct mass scales: it triggers quenching at $\delta_{\rm Q}$, creates the conditions for structural transformation at $\delta_{\rm B}$ by redistributing the halo gas, and loses effectiveness above $M_\mathrm{crit}$. The $\sim 0.5$~dex offset between $\delta_{\rm Q}$ and $\delta_{\rm B}$ in stellar mass traces the halo growth interval over which feedback ramps from onset to peak efficiency. Environment modulates the pathway -- as shown in Section~\ref{subsubsec:q-vs-b}, the ordering of quenching and morphological transformation reverses between field and dense environments -- but the mass thresholds are set by internal processes. We stress that this alignment of mass scales is a coincidence of independently-derived quantities rather than a direct test against our size data; we develop it as a motivated hypothesis for future work (e.g. direct cross-matching against AGN indicators).

\section{Conclusions}
\label{sec:conclusions}

In this work, we combine stellar masses and redshifts inferred using \texttt{pop-cosmos} with COSMOS2025 size/morphology markers to quantify how galaxy size scaling relations depend on morphology across cosmic time. Overall, we show that adding explicit morphological information (via B/T and S\'{e}rsic index) enhances the classical picture of the galaxy size--mass and size--redshift evolution trends, and that using galaxy properties inferred using a population-level forward model provides a coherent connection to information obtained from COSMOS-Web imaging. Our main conclusions are:

\begin{itemize}
    \item \textbf{Star-forming and quiescent galaxies follow distinct functional forms encoding distinct growth physics.} 
    Star-forming galaxies are well described by a single-power law across the full mass range, consistent with continuous in-situ growth regulated by halo angular momentum (Section~\ref{subsubsec:sf-growth}). Quiescent and bulge-dominated galaxies require a double-power law, with a shallow low-mass slope and a steep high-mass slope separated by a pivot mass. This encodes a mass-dependent transition between recently quenched systems and galaxies that have undergone substantial post-quenching size growth, with the same dichotomy imprinted on the size--redshift relation.
    \item \textbf{Intrinsic scatter tracks structure, not star formation, connecting galaxy sizes to halo spin.} 
    The intrinsic scatter of the size--mass relation depends on structural morphology but not on star-formation activity: disc-dominated galaxies show $\sigma_\mathrm{int} \sim 0.13$~dex versus $\sim 0.20$~dex for bulge-dominated systems, while star-forming and quiescent galaxies show similar scatter ($\sim 0.15\text{--}0.16$~dex). This implies that it is the structural configuration of a galaxy -- not its star-formation activity -- that connects to halo occupation (Section~\ref{subsubsec:intrinsic-scatter}). 
    \item \textbf{The coincidence of the three mass scales could link AGN feedback to quenching, structural transformation, and the failure of AGN feedback at the highest masses}. Quenching and structural transformation are partially decoupled, as revealed by the $\sim 0.5$~dex offset between the quiescent pivot mass ($\delta_{\rm Q} = 10.7 \pm 0.2$~dex) and the bulge-dominated pivot mass ($\delta_{\rm B} = 11.1 \pm 0.2$~dex), robust over $0.5 \leq z < 1.5$. The quiescent pivot coincides with the mass scale at which AGN feedback becomes effective and AGN bolometric luminosity peaks in transitioning galaxies. The bulge-dominated pivot coincides with the halo mass at which AGN-driven baryonic redistribution peaks, creating the conditions under which disc heating transforms quenched discs into spheroids (Section~\ref{subsubsec:agn}). At $z < 0.5$, the double-power law pivot is unconstrained, suggesting that in the most massive haloes ($M_{200} \gtrsim 10^{13.5}\,\mathrm{M_\odot}$) AGN heating can no longer prevent gas accretion onto the central galaxy, so quenching is no longer reliably sustained.
    \item \textbf{The three classification schemes capture different stages of the same evolutionary sequence.} 
   The sSFR, B/T, and S\'{e}rsic index classifications are not interchangeable: they produce quantitatively different slopes, intercepts, scatter, and pivot masses (Section~\ref{subsec:interchange}). This is a physical signature -- the three schemes capture different stages of the quenching-then-transformation sequence, and treating them as equivalent obscures the mass-dependent decoupling measured in this work. 
    \item \textbf{Recovering these morphology-dependent trends requires space-based resolution.} The correlation between B/T, stellar mass, and redshift -- and the structural diversity within each sSFR class -- cannot be recovered from ground-based imaging and emerges only in the COSMOS-Web measurements (Section~\ref{subsec:interchange}; Figs.~\ref{fig:ms-z-bt-ssfr-bins} --\ref{fig:confusion-matrix}): ground-based B/T classes agree with the space-based ones only marginally beyond chance, and only 39 per cent of space-based bulge-dominated galaxies are recovered as such. This has direct implications for efforts to learn morphology--property correlations from seeing-limited data.
\end{itemize}

These results demonstrate that galaxy size scaling relations encode not one evolutionary transition but a sequence of physically distinct thresholds, which we hypothesise may be linked to AGN-driven quenching onset, through structural transformation in feedback-reshaped haloes, to the eventual failure of AGN feedback at the highest masses -- recoverable only when star-formation activity and structural morphology are treated as separate dimensions.

\section*{Acknowledgements}

We would like to thank Marko Shuntov, Rohan Varadaraj and Konrad Kuijken for fruitful discussions at various stages of this project. This work has been supported by funding from the European Research Council (ERC) under the European Union's Horizon 2020 research and innovation programmes (grant agreement no.\ 101018897 CosmicExplorer). This work has been enabled by support from the research project grant ‘Understanding the Dynamic Universe’ funded by the Knut and Alice Wallenberg Foundation under Dnr KAW 2018.0067. HVP was additionally supported by the G\"{o}ran Gustafsson Foundation for Research in Natural Sciences and Medicine. This work was performed using resources provided by the Cambridge Service for Data Driven Discovery (CSD3) operated by the University of Cambridge Research Computing Service (\url{www.csd3.cam.ac.uk}), provided by Dell EMC and Intel using Tier-2 funding from the Engineering and Physical Sciences Research Council (capital grant EP/T022159/1), and DiRAC funding from the Science and Technology Facilities Council (\url{www.dirac.ac.uk}).

This study utilises observations collected at the European Southern Observatory under ESO programme ID 179.A-2005 and 198.A-2003 and data products produced by CALET and the Cambridge Astronomy Survey Unit on behalf of the UltraVISTA consortium.

%%%%%%%%%%%%%%%%%%%%%%%%%%%%%%%%%%%%%%%%%%%%%%%%%%
\section*{Data Availability}

The \texttt{pop-cosmos} SPS parameter posteriors described in Section \ref{subsec:pop-cosmos} were originally published in \citet{thorp25b}, and were updated in the present work to v3.0.0, which we have made publicly available on Zenodo \citep{thorp26_mcmc}. The updated v3.0.0 posteriors were run with a higher numerical precision on the \texttt{pop-cosmos} prior, include posterior means and standard deviations as well as percentiles, and have posterior predictions of rest-frame absolute magnitudes in $NUVrJ$. The subset of galaxies used in the present analysis are flagged in v3.0.0 on Zenodo. The COSMOS2025 catalogue published by \citet{shuntov25}, including the morphology measurements described in Section \ref{sec:cosmos2025}, is publicly available via their webpage (\url{https://cosmos2025.iap.fr/catalog.html}). We use the file \verb!COSMOSWeb_mastercatalog_v1.fits! as our input, and include the unique COSMOS-Web IDs of cross-matched galaxies in the updated \citet{thorp26_mcmc} data release. The analysis code is available from the first author on reasonable request.

%%%%%%%%%%%%%%%%%%%%%%%%%%%%%%%%%%%%%%%%%%%%%%%%%%
\section*{Author Contributions}
We outline the different contributions below using keywords based on the Contribution Roles Taxonomy (CRediT; \citealp{brand15}).
\textbf{MNT:} conceptualization; 
formal analysis; methodology; 
software; 
visualization; 
validation; 
investigation; 
data curation; 
writing -- original draft, review \& editing.
\textbf{HVP:} conceptualization; methodology; visualization; 
validation; 
investigation; 
writing -- review \& editing.
\textbf{ST:}
data curation;
methodology; 
software;
validation;
writing -- review \& editing.
\textbf{SD:}
conceptualization;
writing -- review \& editing.
\textbf{DJM:}
methodology; 
writing -- review \& editing.
\textbf{GJ:}
software.
\textbf{AH:}
writing -- review \& editing.
\textbf{BL:}
writing -- review \& editing.
\textbf{BVdB:}
writing -- review \& editing.
\textbf{JL:}
writing -- review \& editing.



\bibliographystyle{mnras}
\bibliography{pop-cosmos} % if your bibtex file is called example.bib

%\begin{thebibliography}{99}
%\end{thebibliography}

%%%%%%%%%%%%%%%%%%%%%%%%%%%%%%%%%%%%%%%%%%%%%%%%%%

%%%%%%%%%%%%%%%%% APPENDICES %%%%%%%%%%%%%%%%%%%%%

\appendix

\section{Completeness and contamination}
\label{sec:completeness_contamination}

\subsection{Typical parameter uncertainties}
\label{sec:uncertainties}

\begin{figure}
    \centering
    \includegraphics[width=\linewidth]{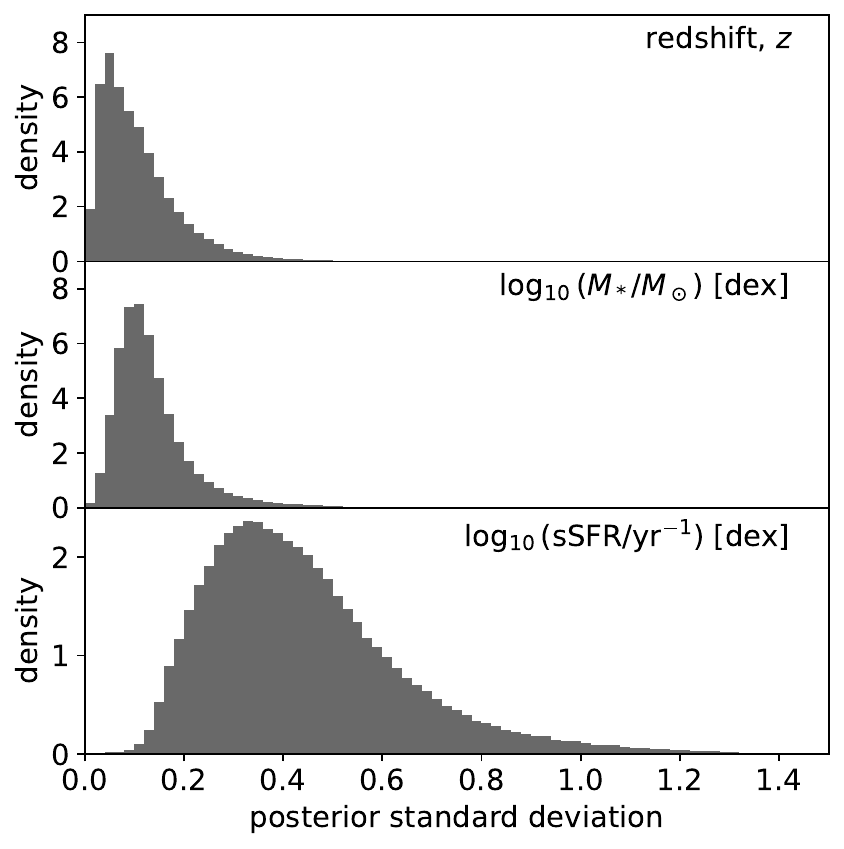}
    \caption{Distribution of parameter uncertainties (posterior standard deviations) for $z$, $\log_{10}(M_*/\mathrm{M}_\odot)$, and $\log_{10}(\mathrm{sSFR}/\mathrm{yr}^{-1})$ from \texttt{pop-cosmos} for all mass-complete COSMOS2020 galaxies with $z\leq3.0$  ($308\,660$ galaxies).}
    \label{fig:stddevs}
\end{figure}

In Figure \ref{fig:stddevs}, we show the typical uncertainties (posterior standard deviations) for the SPS parameter estimates we take from the \citet{thorp25b} \texttt{pop-cosmos} catalogue. For $99.4$~per~cent of this sample, the posterior standard deviation on $z$ is below $0.5$ (cf.\ the requirement set by \citealp{carreira26}). Similarly, $99.0$~per cent of the sample has a stellar mass uncertainty smaller than $0.5$~dex. The posteriors on $\log_{10}(\mathrm{sSFR}/\mathrm{yr}^{-1})$ tend to be broader, with $31.9$~per cent of galaxies having an uncertainty larger than $0.5$~dex. However, the majority of cases ($97.9$~per cent) are bounded below 1.0~dex. We investigate the implications of sSFR uncertainty further in Section \ref{sec:ssfr_reliability}, finding that the vast majority of galaxies have a secure sSFR-based classification into star-forming/quiescent samples.

\begin{figure}
    \centering
    \includegraphics[width=\linewidth]{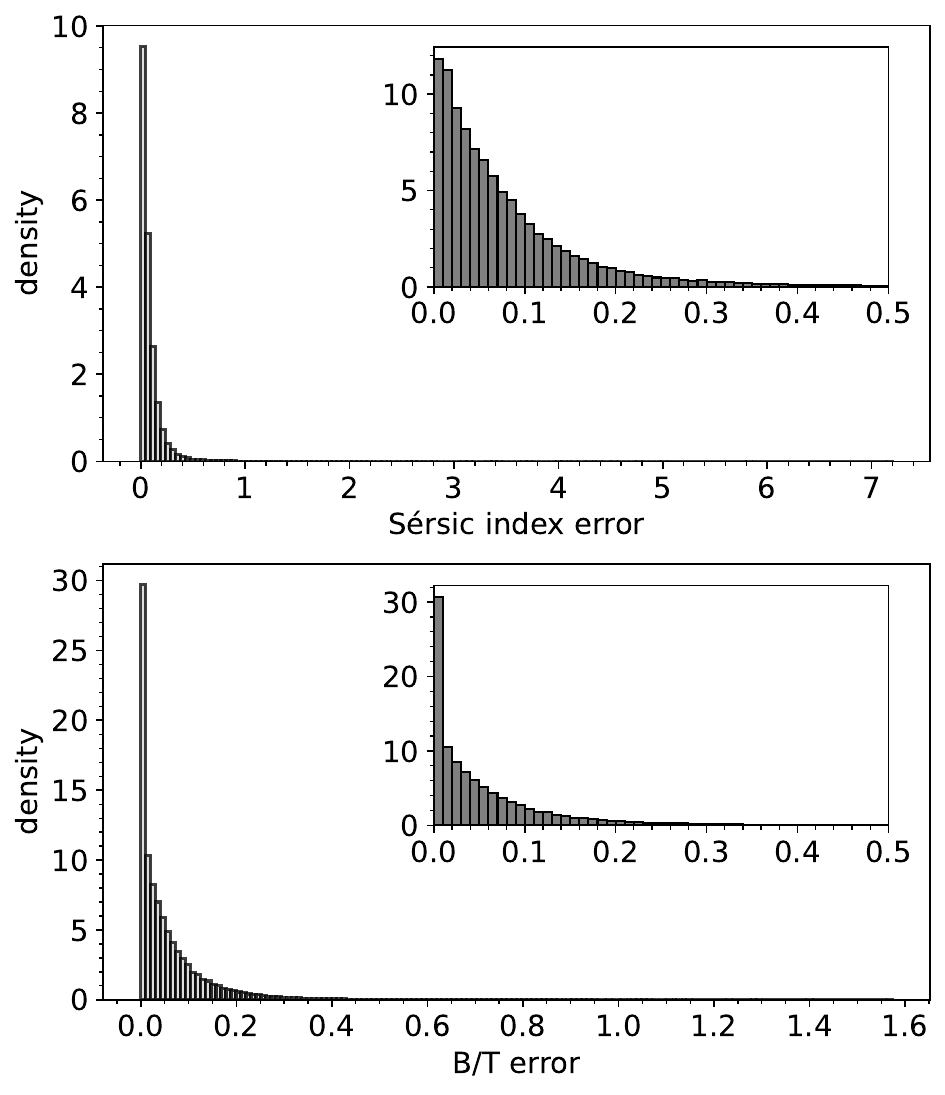}
    \caption{Distribution of parameter uncertainties for S\'{e}rsic index $n$ (\textit{top}) and B/T (\textit{bottom}) from \citet{shuntov25} for the crossmatched galaxies with $z \leq 3.0$ ($99\,369$ galaxies). The inset graph in each panel shows a zoom-in of the values of the error for each parameter, going between $0.0$ and $0.5$.}
    \label{fig:stddevs-morpho}
\end{figure}

Figure \ref{fig:stddevs-morpho} shows the typical uncertainties of the B/T (in F115W for galaxies $z < 2.0$ and F150W for galaxies $2.0 \leq z \leq 3.0$) measurements, as well as the S\'{e}rsic index measurements, taken from \cite{shuntov25}. Both distributions are strongly peaked towards small values. The S\'{e}rsic index has a median uncertainty of $0.06$, with $99.0$~per cent of the sample having the uncertainty below $0.5$. The B/T uncertainties are even more tightly concentrated, with a median of $0.03$ and $81.1$~per cent of the sample having uncertainties smaller than $0.1$. These uncertainties are small relative to the widths of the bins used for our classifications ($n = 2.5$ for the late-/early-type split; B/T $= 0.2$ and $0.6$ for the disc/bulge selection), such that the majority of galaxies lie many standard deviations from their nearest dividing line. We also investigate the implications of these uncertainties further in Section \ref{sec:ssfr_reliability}, finding that the vast majority of galaxies have a secure morphological classification.

\subsection{Reliability of sample splits}
\label{sec:ssfr_reliability}
\begin{figure}
    \centering
    \includegraphics[width=\linewidth]{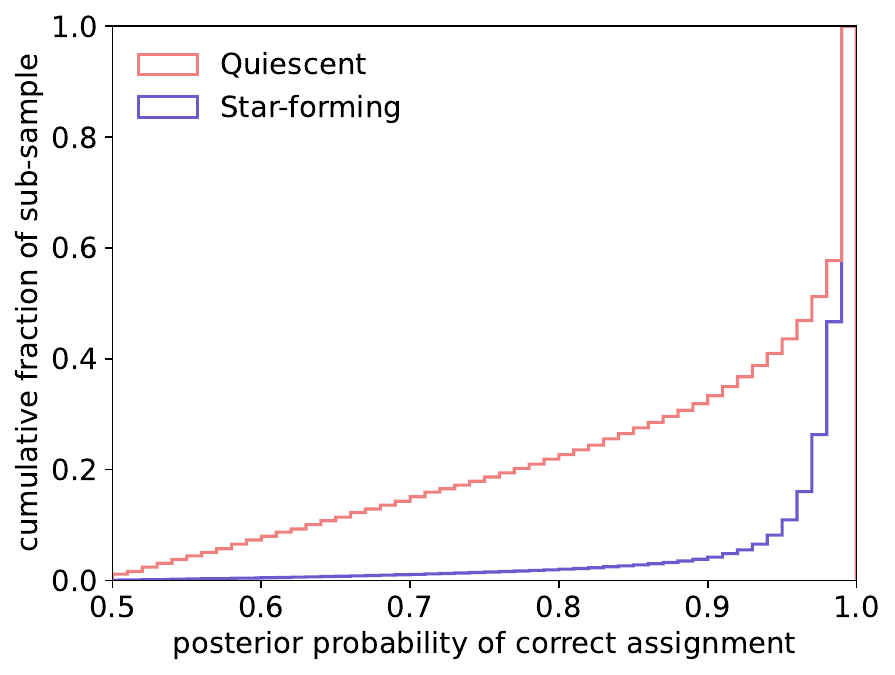}
    \caption{Posterior belief in the division of all mass-complete $z\leq3.0$ COSMOS2020 galaxies from \citet{thorp25b} into star-forming and quiescent populations. For each sub-sample of galaxies, identified as star-forming (blue) or quiescent (red) based on their posterior medians, we plot the fraction of cases where the posterior probability of the assignment is less than or equal to a certain level. Fractions are normalised relative to the total in each sub-sample ($14\,406$ quiescent; $294\,254$ star-forming).}
    \label{fig:p_correct_ssfr}
\end{figure}

We assign galaxies into star-forming and quiescent sub-samples based on their posterior median sSFR being above or below $10^{-11}\,\mathrm{yr}^{-1}$. We now investigate whether this procedure induces Eddington bias, demonstrating that this has negligible impact. Eddington bias could potentially arise if galaxies from the smaller quiescent population are scattered into the larger star-forming sub-sample. To quantify this, for each $z\leq3.0$ COSMOS2020 galaxy in the two sub-samples that is above our mass completeness limit, we compute the posterior probability that its star-forming/quiescent assignment is correct, and plot the cumulative distribution of these probabilities in Figure \ref{fig:p_correct_ssfr}. The majority of the star-forming sample is high-confidence -- $92$~per cent of galaxies in this sample are assigned to it with $>95$~per cent posterior probability -- whilst the number of low-confidence galaxies in the quiescent sample is small -- only $\sim18$~per cent of these have posterior probability below the $75$~per cent level. We thus conclude that our posterior median-based assignment of galaxies into sub-populations is robust.

\begin{figure}
    \centering
    \includegraphics[width=\linewidth]{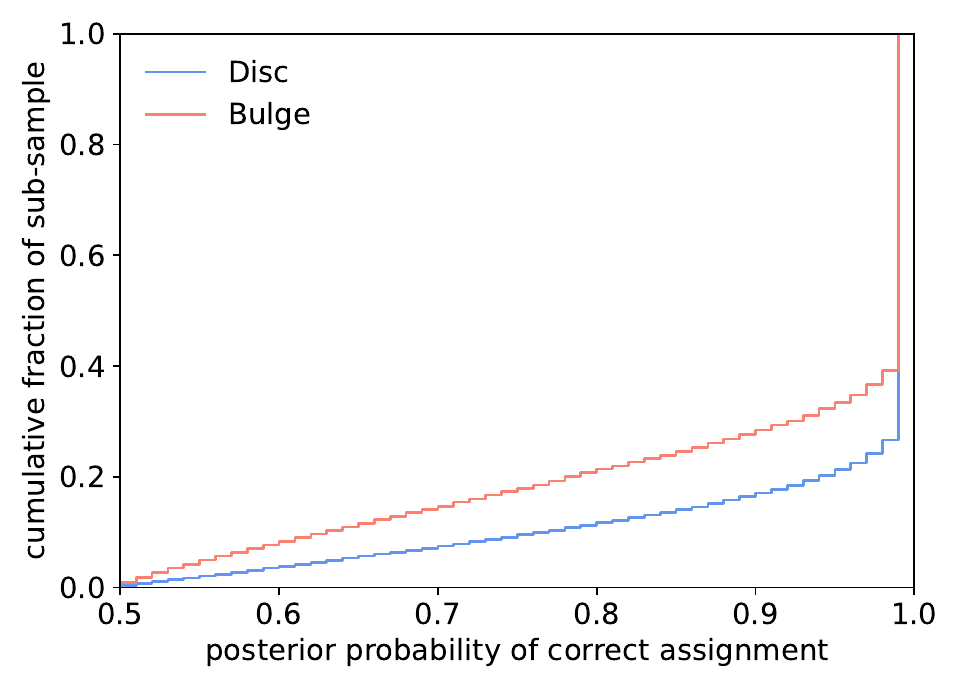}
    \caption{Reliability of the disc/bulge classification. For each galaxy, assigned to the disc-dominated (B/T $\leq 0.2$, blue) or bulge-dominated (B/T $\geq 0.6$, red) class from its measured B/T, we compute the probability that a draw from a Gaussian centred on the measured B/T falls in the assigned class, and plot the fraction of each sub-sample with a probability of correct assignment less than or equal to a given level. The intermediate population ($0.2 <$ B/T $< 0.6$) is excluded. Fractions are normalised relative to the total in each sub-sample ($66\,704$ disc-dominated; $9\,877$ bulge-dominated).}
    \label{fig:p_correct_bt}
\end{figure}

\begin{figure}
    \centering
    \includegraphics[width=\linewidth]{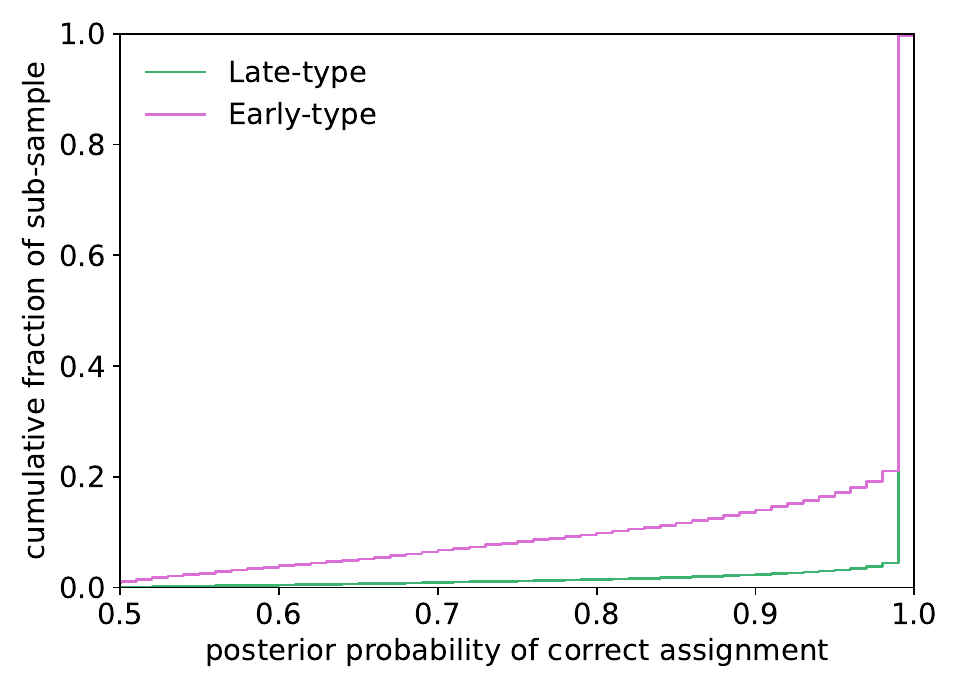}
    \caption{Reliability of the late/early-type classification. For each galaxy, assigned to the late-type ($n \leq 2.5$, green) or early-type ($n > 2.5$, pink) class from its measured $n$, we compute the probability that a draw from a Gaussian centred on the measured $n$ falls in the assigned class, and plot the fraction of each sub-sample with a probability of correct assignment less than or equal to a given level. Fractions are normalised relative to the total in each sub-sample ($88\,222$ late-type; $11\,147$ early-type).}
    \label{fig:p_correct_n}
\end{figure}

We repeat this test for the two morphological classifications. While \texttt{pop-cosmos} posteriors exist for the sSFR case, the COSMOS2025 catalogue does not provide posteriors for the structural parameters, so we approximate each galaxy's B/T (or S\'{e}rsic-index) posterior as a Gaussian centred on the \texttt{SourceXtractor++} value with width equal to its quoted uncertainty \citep{shuntov25}, and compute the probability that each galaxy's B/T falls in the assigned class. For the disc-dominated (B/T $\leq 0.2$) and bulge-dominated (B/T $\geq 0.6$) split, we compute the posterior probability that each galaxy is placed in its assigned class and plot the cumulative distributions in Figure~\ref{fig:p_correct_bt}. The disc-dominated sample is assigned with high confidence, with $83.6$~per cent of galaxies having a probability of correct assignment greater than $90$~per cent. The bulge-dominated sample is somewhat less confidently assigned but remains secure, with $72.4$~per cent of galaxies assigned at greater than $90$ per cent probability. Similarly, for the late-type ($n \leq 2.5$) and early-type ($n > 2.5$) S\'{e}rsic-index split, we compute the probability of correct assignment using the S\'{e}rsic-index uncertainties, shown in Figure \ref{fig:p_correct_n}. Here the assignment is even more secure: $97.7$~per cent of late-type and $86.2$~per cent of early-type galaxies are assigned at greater than $90$~per cent probability. As with the sSFR classification, the small typical uncertainties on B/T and $n$ relative to the bin widths mean that the overwhelming majority of galaxies sit far from their nearest dividing line, and we conclude that the morphological assignments are robust.

\subsection{Crossmatching failures between COSMOS catalogues}
\label{sec:completeness}

\begin{figure}
    \centering
    \includegraphics[width=\linewidth]{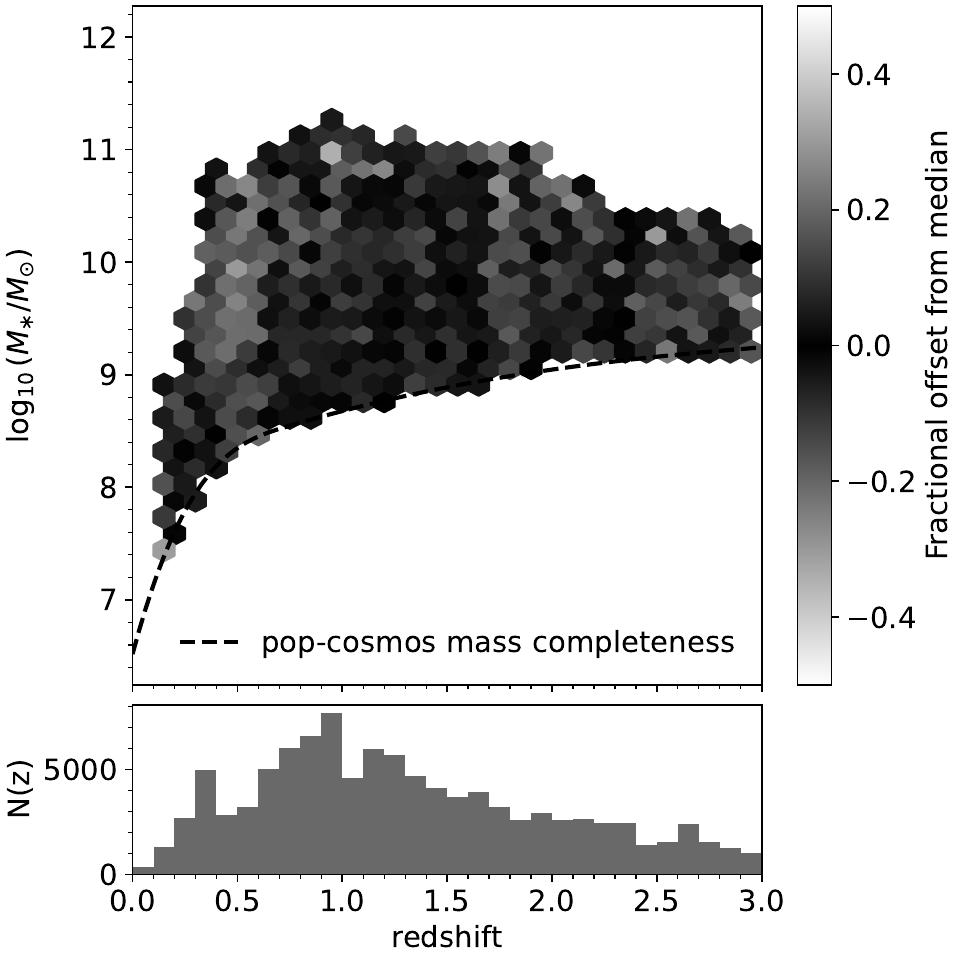}
    \caption{Stellar mass versus redshift for the COSMOS2025 crossmatched sample described in Section \ref{sec:cross-match} \textit{(top panel}), along with its redshift distribution \textit{(bottom panel)}. The black dashed line shows the mass completeness limit from \citet{thorp25b}. The colour scale shows the fractional deviation from the median crossmatch ratio in each mass-redshift bin, where the ratio is defined as the number of crossmatched COSMOS2025 sources divided by the area-corrected number of parent COSMOS2020 sources.}
    \label{fig:completeness}
\end{figure}

\begin{figure*}
    \centering
    \includegraphics[width=\linewidth]{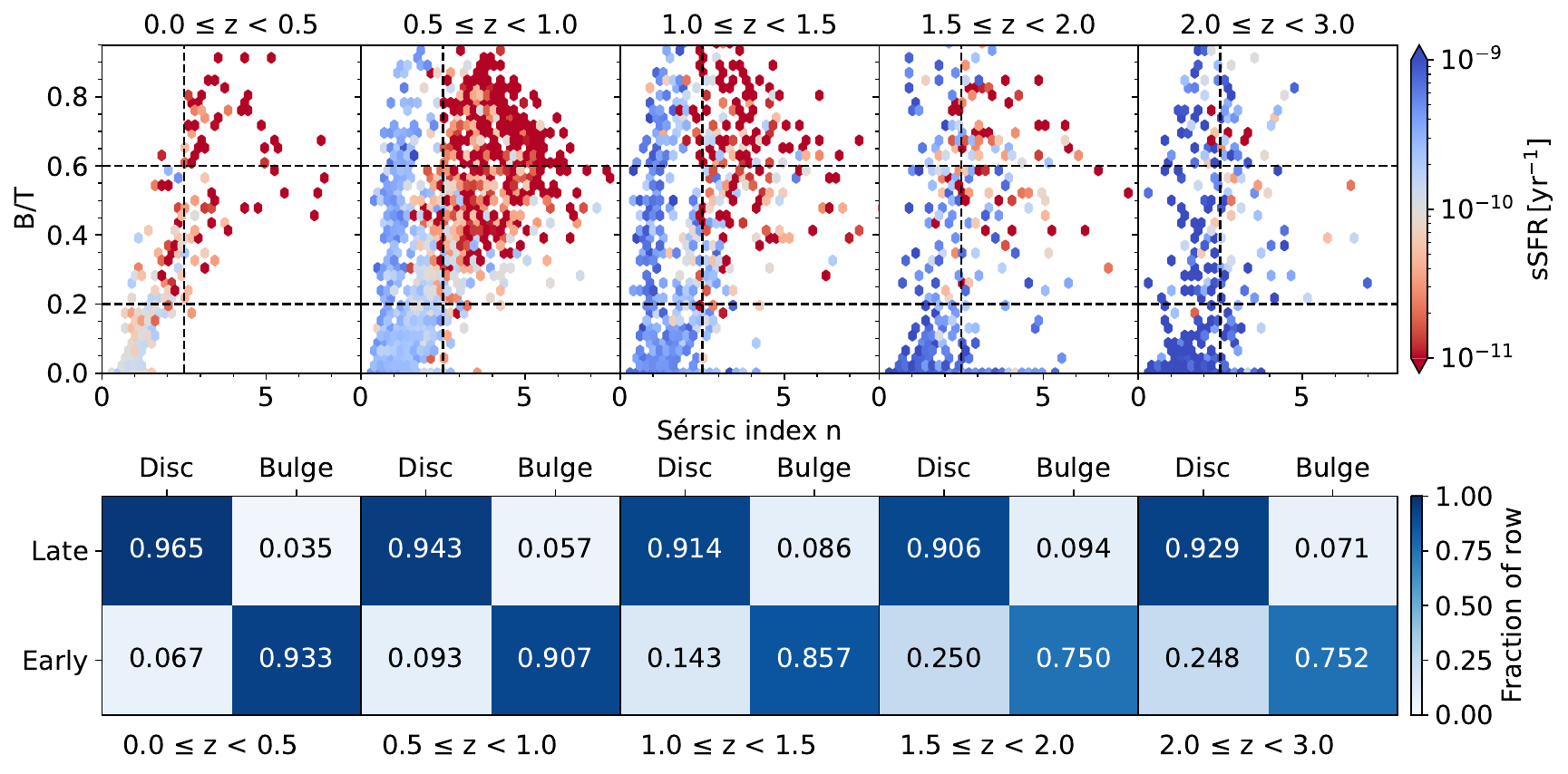}
    \caption{S\'{e}rsic index $n$ versus bulge-to-total ratio B/T, colour-coded by sSFR, for the high-mass sample ($\log_{10}(M_{\ast}/\mathrm{M}_\odot) > 10.3$) in five redshift bins. The dashed lines mark the classification thresholds $n = 2.5$ (late/early-type split) and $\mathrm{B/T} = 0.2, 0.6$ (disc/intermediate/bulge-dominated split), calculated from the F115W filter for galaxies at $0.0 \leq z < 2.0$ and from the F150W filter for galaxies at $2.0 \leq z \leq 3.0$. The correlation between $n$ and B/T is positive but broad at all redshifts, and degrades substantially above $z \sim 1.5$, where early-type galaxies ($n > 2.5$) are increasingly spread across the full B/T range \textit{(top panel)}. Contamination fraction matrix showing the fraction of late-type ($n \leq 2.5$) and early-type ($n > 2.5$) galaxies that are disc-dominated ($\mathrm{B/T} \leq 0.2$) or bulge-dominated ($\mathrm{B/T} \geq 0.6$) in each redshift bin for the full sample. The fraction of early-type galaxies that are disc-dominated rises from $6.7$~per cent at $0.0 \leq z < 0.5$ to $25$~per cent at $1.5 \leq z < 2.0$, demonstrating that the $n > 2.5$ cut increasingly selects a structurally heterogeneous population at higher redshift and explaining why no transition mass $\delta$ is recovered for the early-type fits \textit{(bottom panel)}.}
    \label{fig:bt-vs-n}
\end{figure*}

Our fiducial sample is based on crossmatching sources between COSMOS2025 \citep{shuntov25} and COSMOS2020 \citep{weaver22} within a $0.5''$ radius (see Section~\ref{sec:cross-match}). To ensure that no systematics are introduced by this step, we must verify that failed crossmatches occur `at random', without dependence on galaxy properties, i.e., that the crossmatched sample is representative of the parent COSMOS2020 catalogue from which the galaxy-properties are derived using \texttt{pop-cosmos}. 
To quantify this, we compute the ratio of the number of crossmatched COSMOS2025 galaxies to the number of mass-complete COSMOS2020 galaxies, where the latter is rescaled by the ratio of the two survey areas to account for their differing footprints. This ratio is computed in bins of $0.1$~dex in stellar mass and $0.1$ in redshift across the mass-redshift plane, restricted to bins containing at least $100$ COSMOS2020 galaxies to ensure statistical reliability. To isolate any systematic trends, we normalise each bin's ratio by the median ratio across the full plane, such that the resulting quantity reflects the fractional deviation of each bin's crossmatch completeness from the global median. As shown in Figure \ref{fig:completeness}, the normalised ratio is consistent with zero across the mass-redshift plane above the \cite{thorp25b} mass completeness limit, with no coherent gradients in either stellar mass or redshift. This confirms that the crossmatched sample does not suffer from mass- or redshift-dependent selection effects beyond those already present in the parent COSMOS2020 catalogue, and that the remaining sample is fully representative of it.

\subsection{Non-equivalence of morphology markers}
\label{bt-vs-n}

The S\'{e}rsic index $n$ and the bulge-to-total ratio B/T are related but distinct structural quantities: B/T is measured from a two-component bulge+disc decomposition and captures the fractional contribution of the bulge to the total luminosity, making it a physically motivated quantity for separating galaxies with structurally dominant bulges from those dominated by their discs, whereas $n$ is derived from a single-component fit to the full composite light profile and therefore reflects an average over bulge and disc contributions rather than isolating either (see Section \ref{sec:cosmos2025}). As \cite{gadotti09} demonstrated on a large low-redshift sample, and as confirmed for the COSMOS-Web population by \cite{yang26}, the correlation between $n$ and B/T is positive but broad: galaxies with similar B/T span a wide range of $n$, and vice-versa, such that a threshold cut at $n > 2.5$ selects a more heterogeneous population than a B/T-based classification, mixing classical bulges, pseudo-bulges, and disc-dominated systems whose composite profiles happen to be centrally concentrated. This scatter is compounded at the redshifts probed here by two additional effects. First, the \textsc{SourceXtractor++} fits in the COSMOS2025 morphological catalogue are performed jointly across all four NIRCam bands, so the effective rest-frame wavelength shifts progressively blueward with redshift. Second, since S\'{e}rsic index is wavelength-dependent, with bulge-dominated profiles appearing more concentrated in redder rest-frame light \citep{yang26}, the joint NIRCam fit could underestimate the concentration of bulge-dominated systems at high redshift, where the effective rest-frame wavelength is bluer.

Both effects come into play when we fit for the transition mass $\delta$. The top panel of Fig.~\ref{fig:bt-vs-n} shows the $n$--B/T relation for our high-mass sample ($m_{\ast} >  10.3$), to contrast it with the first two panels of Fig.~7 in \citet{yang26}. Unlike the full COSMOS-Web population studied by \citet{yang26}, where the median B/T flattens above $n \sim 2.5$, our sample retains a positive correlation throughout the full range of $n$. Nevertheless, the scatter remains substantial, and the correlation degrades with increasing redshift. The bulge-dominated sample (B/T $\geq0.6$) is a stringent structural selection that excludes the large intermediate population ($0.2 < B/T < 0.6$), maintaining the homogeneity required for the double-power law to reveal a transition in the size--mass relation in at least some redshift bins. The early-type ($n > 2.5$) sample, by contrast, draws substantially from this intermediate B/T regime. In the bottom panel of Fig.~\ref{fig:bt-vs-n} -- particularly at high redshift -- the correlation between S\'{e}rsic index and B/T breaks down and the $n > 2.5$ cut no longer reliably identifies genuinely bulge-dominated systems, with the fraction of early-type galaxies that are disc-dominated rising from 6.7~per cent at $0.0 \leq z < 0.5$ to 25~per cent at $1.5 \leq z < 2.0$. The resulting structural diversity washes out any transition between distinct size–mass regimes, rendering $\delta$ unconstrained across all early-type redshift bins.

%%%%%%%%%%%%%%%%%%%%%%%%%%%%%%%%%%%%%%%%%%%%%%%%%%

% Don't change these lines
\bsp	% typesetting comment
\label{lastpage}
\end{document}